\documentclass[
  aps,            
  pre,            
  reprint,        
  nofootinbib,    
  floatfix,
  superscriptaddress,
  amsmath,amssymb 
]{revtex4-2}

\usepackage{graphicx}
\usepackage{xcolor}
\usepackage{hyperref}
\usepackage{enumitem}
\usepackage{mathtools}
\usepackage{microtype}

 \newcommand{\edit}[1]{#1} 

\graphicspath{{figures/}}

\begin{document}

\title{Emergent Topology of Optimal Networks for Synchrony}

\author{Guram Mikaberidze}
\affiliation{School of Computing, University of Wyoming}

\author{Dane Taylor}
\affiliation{School of Computing, University of Wyoming}
\affiliation{Department of Mathematics \& Statistics, University of Wyoming}

\date{\today}

\begin{abstract}
Designing high-performing networks requires optimizing for functionality while respecting physical, spatial, or budget constraints. Yet, mathematical and computational tools to design such systems remain limited, particularly for collective dynamics arising from heterogeneous dynamical units. Here, we develop a gradient-based optimization framework to identify synchrony-optimal weighted networks under a constrained coupling budget. The resulting networks exhibit counterintuitive properties: they are sparse, bipartite, elongated, and extremely monophilic (i.e., the neighbors of any node are similar to one another while differing from the node itself). These structural patterns persist across dynamical models ranging from the power-grid swing equations to chaotic Rössler systems, suggesting broad applicability to coupled oscillator technologies. To gain insight, we develop a ``constructive'' theory for coupled Kuramoto oscillators: a nonlinear differential equation identifies which pairs of nodes are coupled, while a variational principle prescribes the budget allocated to each node. Dynamics unfolding over optimal networks provably lack a synchronization threshold; instead, as the budget exceeds a calculable critical value, the system globally phase-locks, exhibiting critical scaling at the transition. Together, our findings offer design principles for synchrony-dependent technologies with potential applications ranging from microgrids to laser arrays and quantum oscillators.
\end{abstract}

\maketitle

\section*{Introduction}

Synchronization of coupled oscillators underlies collective behavior across a remarkable range of natural and engineered systems, from cortical rhythms and circadian pacemakers to electrical power grids, coupled laser arrays, and Josephson junction circuits \cite{kuramoto1975international,pikovsky2001universal}. The existence of strong coherence in such systems
depends sensitively on the structure of the network linking   oscillators \cite{arenas2008synchronization, rodrigues2016kuramoto}. Decades of research have shown that even small structural modifications can dramatically enhance or impair synchrony \cite{watts1998collective, donetti2005entangled, schafer2022understanding}, motivating a sustained search for the topological features that best promote synchronization \cite{pecora1998master, barahona2002synchronization, nishikawa2003heterogeneity, nishikawa2006synchronization,nishikawa2010network, brede2008synchrony, pinto2015optimal,buzna2009synchronization,brede2018competitive, ye2025optimal, sorrentino2008adaptive,kelly2011topology,skardal2014optimal, taylor2016synchronization, skardal2016optimal, skardal2017optimal, skardal2019synchronization, chamlagai2022grass}. 
%
%
In the context of synchronization for identical dynamical units, master stability function theory \cite{pecora1998master} has led to insights about the types of structures that enhance synchronizability \cite{barahona2002synchronization,nishikawa2003heterogeneity} as well as the development of ``constructive theory''---that is, a direct characterization of network topologies that best optimize the synchronization of identical dynamical units \cite{nishikawa2006synchronization,nishikawa2010network}.


Real-world systems, however, often contain heterogeneity, and it is well known that synchrony optimization for identical and non-identical dynamics can give rise to vastly differing network structures. 
%
Related work includes computational studies using
evolutionary~\cite{brede2008synchrony, brede2018competitive, ye2025optimal} and heuristic algorithms~\cite{sorrentino2008adaptive,kelly2011topology}; however, these have yielded inconsistent structures in the absence of a unifying theory. 
%
Optimization theory for networks of heterogeneous oscillators has centered on the synchrony alignment function (SAF)~\cite{skardal2014optimal, taylor2016synchronization, skardal2016optimal,taylor2016synchronization, skardal2017optimal, skardal2019synchronization, chamlagai2022grass}, which measures synchronizability by considering how oscillator heterogeneity  aligns with network structure. While the SAF simplifies the quantification and numerical optimization of network structures for synchrony, it does not provide a direct constructive characterization of the networks resulting after optimization. It also relies on a linear approximation for the limit of strong synchrony.
Thus, 
existing knowledge about the network topologies that best synchronize heterogeneous dynamical units remains limited.

%

\begin{figure*}[!t]
    \centering
    \includegraphics[width=\textwidth]{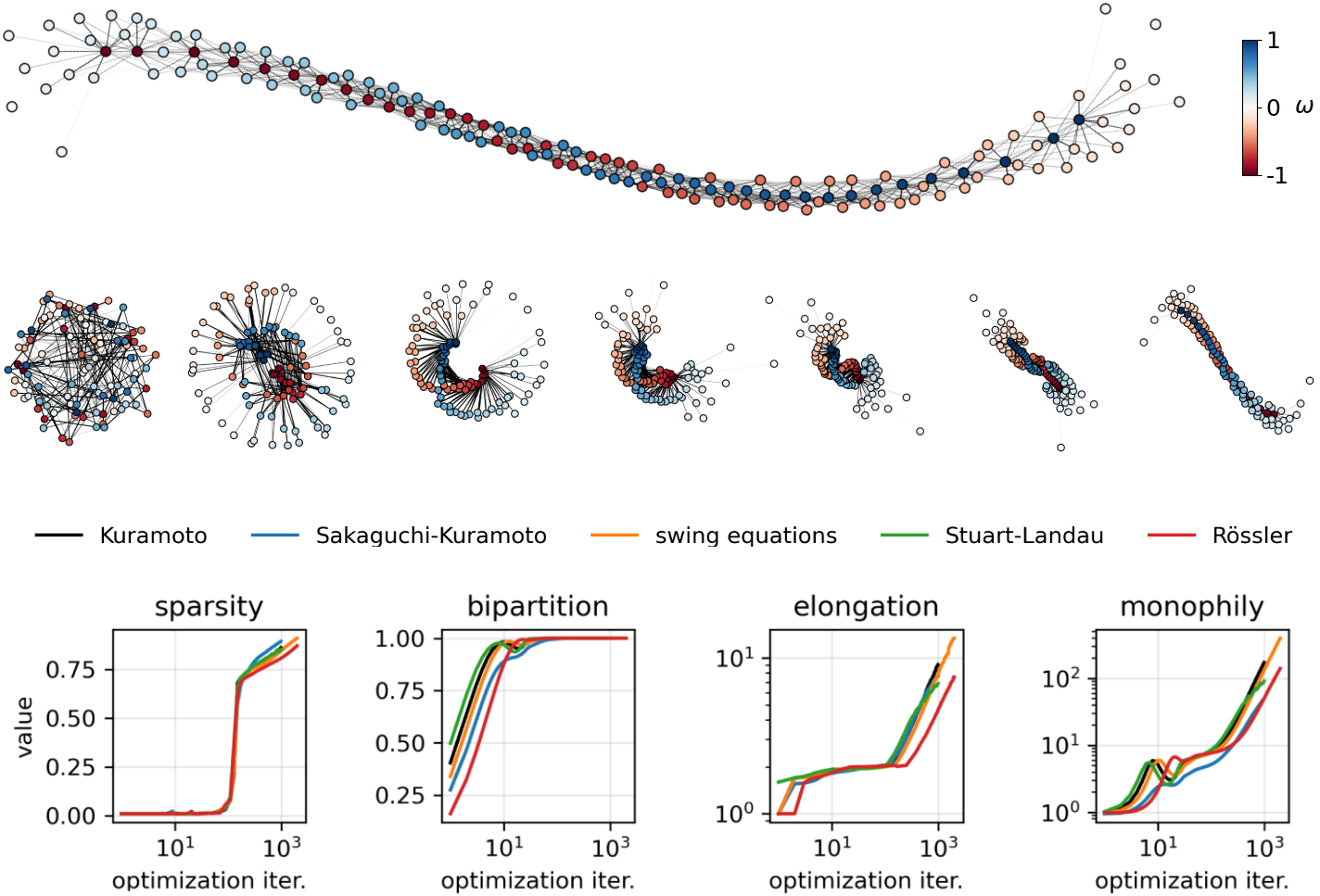}
    \caption{
    \textbf{Network optimization for synchrony under a limited coupling budget $b$.}
        A representative network is shown on \textbf{top}, while snapshots at successive stages of optimization are visualized in the \textbf{middle row}.
        Node colors encode intrinsic frequencies $\omega_i$; edge thicknesses reflect edge weights $A_{ij}>0$. 
        The emergent structure is highly sparse and bipartite, with edges exclusively linking nodes of contrasting frequencies (e.g., blue to red). It is surprisingly elongated, with long average path lengths. The network also exhibits extremely strong monophily: the neighbors of any node are strikingly homogeneous. This effect is readily observed at the periphery. For example, on the left, dark-red nodes have neighbors that are all shaded light-blue.  
        \textbf{Bottom row:} Evolution of four topological metrics during optimization of networks for five paradigmatic synchronizing systems: The Kuramoto model, Sakaguchi–Kuramoto model with phase frustration, power grid swing equations, Stuart–Landau phase-amplitude oscillators, and chaotic R\"ossler oscillators (details in Methods section). \textit{Sparsity} is measured by the fraction of negligibly small edge weights ($A_{ij}\le 10^{-8}$). \textit{Bipartition} is measured as  $1 - \tfrac{1}{2}{\sum_i \left| \lambda_i + \lambda_{N+1-i} \right|} \ /\ {\sum_i \left| \lambda_i \right|}$, where $\{\lambda_i\}$ is the adjacency matrix spectrum, which must be symmetric for bipartite networks. \textit{Elongation} is measured by the average edge count across all pair-wise shortest paths (with path-lengths computed using edge lengths $l_{ij}=1/A_{ij}$). \textit{Monophily} is quantified here as $\tfrac{1}{N^2}\sum_{ij}(\omega_i-\omega_j)^2 \ /\ \sum_{ij}P_{ij}(\omega_i-\omega_j)^2$, the ratio of unweighted to two-hop-weighted pairwise frequency variance, where $P_{ij}=(\boldsymbol{A}^2)_{ij}/\sum_{kl}(\boldsymbol{A}^2)_{kl}$ weights node pairs by their two-hop coupling.
    }
    \label{fig:network}
\end{figure*}

Herein, we develop computational and analytical tools to obtain and characterize network topologies that optimize synchronization among heterogeneous oscillators.
Computationally,  we take advantage of  hardware and software for machine learning \cite{paszke2017automatic,torchdiffeq} to develop robust, gradient-based optimization algorithms that uncover highly efficient networks with counterintuitive structure (see Fig.~\ref{fig:network}). Similar structures are recovered across diverse models
including the Kuramoto model, Sakaguchi-Kuramoto model, swing equations, Stuart-Landau oscillators, and chaotic R\"ossler oscillators (Fig.~\ref{fig:network}, bottom row).
Optimal networks uncovered by our framework overturn long-standing assumptions. We find that sparsity, bipartition, elongated structure, and monophily---not homogeneity or small-world topology---enable optimal synchrony under constraints. The first three are familiar structural descriptors; monophily is less so. It refers to a network in which a node's neighbors are similar to one another in some attribute but differ from the node itself, contrasting homophily, whereby nodes connect to others similar to themselves.


We find that these structural patterns are emergent structural phenomena for networks that are optimized while being constrained to a fixed \textit{coupling budget}---a limit on the summation of edge weights. That is,
we reframe synchronization as a constrained optimization problem in which both the topology and edge weights of a network are designed to maximize synchrony subject to a limited coupling budget,
which for the Kuramoto model, replaces the global coupling constant.
Our motivation for this reframing is that
real networks usually operate within stringent physical, economic, or spatial constraints. Upgrading a transmission line in a power grid, for example, draws from a finite resource pool, limiting improvements elsewhere. Similarly, in neuromorphic and spintronic systems, only a subset of possible couplings can be realized due to limits on heat dissipation, parasitic effects, or spatial layout.  In such contexts, synchronization cannot be tuned by simply scaling a global parameter; instead, it must emerge from the strategic allocation of a limited coupling budget across the network. These constraints are not mere technicalities; they fundamentally shape network architecture. Despite its broad relevance, from critical infrastructure to nanoscale devices, the problem of achieving optimal synchrony under explicit coupling constraints remains analytically unresolved and algorithmically challenging.

To obtain theoretical insight into why these synchrony-optimizing structural patterns emerge, we develop constructive theory for a general family of coupled oscillators (see Eq.~\eqref{oscillator_eq}). We derive an optimal pairing function that identifies which pairs of nodes should be coupled by edges. This yields two locally optimal network configurations, each recapitulating different anecdotal observations resulting from prior network studies \cite{brede2008synchrony, brede2018competitive, ye2025optimal,sorrentino2008adaptive,kelly2011topology,skardal2014optimal, skardal2016optimal, skardal2017optimal, skardal2019synchronization}.
Our analysis enables a direct comparison between these two solutions, identifies one as the global optimum, and unifies this literature under a common theory. For the Kuramoto model, we also analytically prescribe the optimal allocation of node strengths (i.e., weighted node degrees) under resource constraints, operating directly in the fully nonlinear interaction regime. 
%
The framework reconciles previous anecdotal observations, yields testable predictions, and offers practical guidance for the design of high-performance networks in domains where synchronization is essential but resources are inherently limited.

\section*{Results}
\subsection*{Gradient-based optimization framework}

We consider a network of $N$ heterogeneous oscillators with frequencies $\{\omega_i\}_{i=1}^N$. The elements of the weighted adjacency matrix $\boldsymbol{A}$ serve as parameters to be optimized subject to a constrained \textit{coupling budget}
\begin{equation}\label{budget}
b = \frac{1}{N} \sum_{ij} A_{ij},
\end{equation}
which represents the average coupling budget per node. Given $b$, we seek to construct and study synchrony-optimizing weighted networks having a total budget $N b$. 

To facilitate gradient-based network optimization, we encode $\boldsymbol A$ using a matrix of tunable parameters $\boldsymbol{P} \in \mathbb{R}^{N\times N}$ as
\begin{equation}\label{adj_from_P}
    A_{ij} = N b \frac{P_{ij}^2 + P_{ji}^2}{2\sum_{kl} P_{kl}^2} (1-\delta_{ij}).
\end{equation}
Here $\delta_{ij}$ is the Kronecker delta and the term $(1-\delta_{ij})$ masks the diagonal elements. This is a smooth map from arbitrary $\boldsymbol P$ to symmetric, non-negative adjacency matrices that adhere to the budget constraint.

Given some dynamical process unfolding on this network $d\boldsymbol{\theta}/dt=\boldsymbol{f}(\boldsymbol{\theta},\boldsymbol{A}, t)$ we define a scalar measure of synchrony $r(t)=r(\boldsymbol\theta(t))$. Focusing on the time average $\langle r\rangle$ of $r(t)$ over some duration $t\in\mathcal{T}$, we seek to obtain the optimal parameter matrix $\boldsymbol{P}$ that maximizes $\langle r \rangle$.
\begin{figure*}[!t]
    \centering
    \includegraphics[width=1\textwidth]{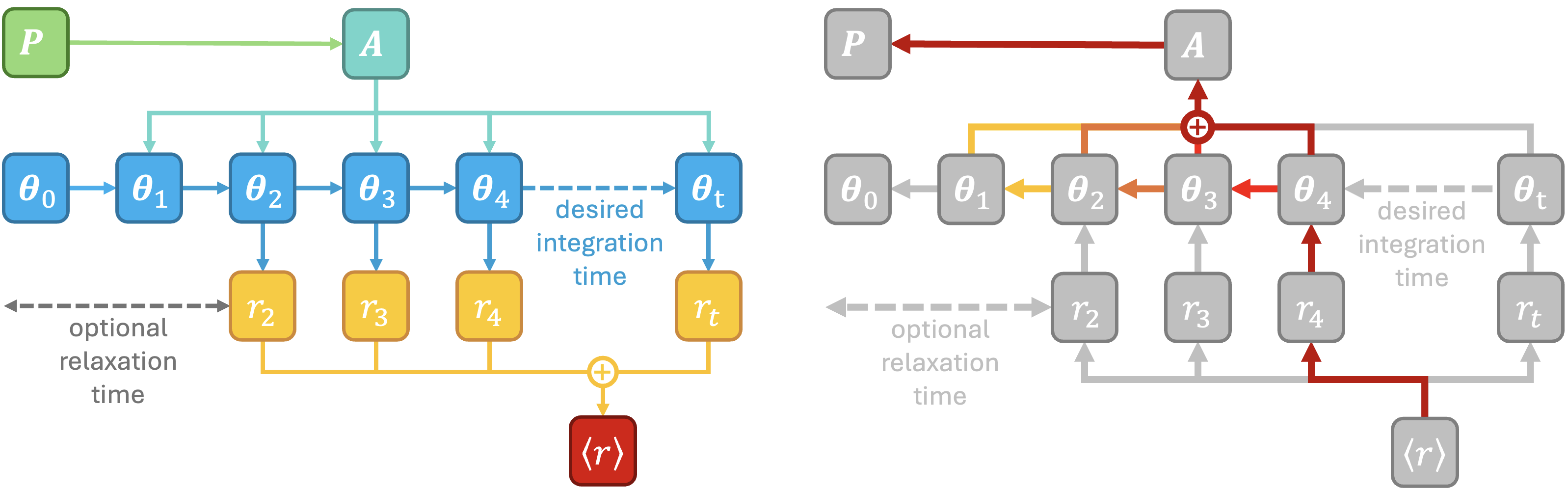}
    \caption{
        \textbf{Computational graph for gradient-based network optimization.}  
        \textbf{Left:} Forward pass. A tunable parameter matrix $\boldsymbol{P}$ defines a symmetric adjacency matrix $\boldsymbol{A}$ obeying the budget constraint. The dynamical system $d\boldsymbol{\theta}/dt=\boldsymbol{f}(\boldsymbol{\theta},\boldsymbol{A}, t)$ is then numerically integrated from an initial condition $\boldsymbol{\theta}_0$, where each subsequent state $\boldsymbol{\theta}_t$ depends on the previous state $\boldsymbol{\theta}_{t-1}$ and on $\boldsymbol 
        A$. We compute  $r_t$ at each time step and then the objective function $\langle r \rangle$ over a duration $\mathcal{T}$. An initial time window may be discarded to ignore the initial transient behavior.  
        \textbf{Right:} Backward pass. The dependency structure of $d\langle r \rangle / dP_{ij}$ is illustrated via so-called ``backpropagation''---that is, the chain rule of calculus. 
        For clarity, we use color to highlight only the backward pathways passing through $r_4$, which correspond explicitly to the term $\frac{\partial \langle r \rangle}{\partial r_4}\frac{d r_4}{d P_{ij}}$. 
        The computational graph and chain-rule are automatically tracked and computed using standard computational techniques for machine learning  \cite{paszke2017automatic,torchdiffeq}. The code is openly available on \textsc{GitLab} \cite{mikaberidze2025networkoptimization} and has been further developed into a customizable Python package \cite{mikaberidze2026gradnet}.
    }
    \label{fig:computational_graph}
\end{figure*}

As illustrated in Fig.~\ref{fig:computational_graph}, numerical integration can be characterized as  a fully differentiable computational graph in which the parameters $\boldsymbol{P}$ determine the adjacency matrix $\boldsymbol{A}$, which governs the oscillator dynamics and yields a scalar synchrony measure for optimization: $\langle r\rangle$. We obtain the optimal parameter matrix $\boldsymbol{P}$ that maximizes $\langle r \rangle$ by following the gradient, and the gradients of $\langle r \rangle$ with respect to $P_{ij}$ are obtained using the automatic differentiation capabilities of \textsc{PyTorch} and \textsc{torchdiffeq} \cite{paszke2017automatic,torchdiffeq} (similar to \cite{ricci2021kuranet}). See the Methods section for further computational details.

\subsection*{Emergence of a sparse, bipartite, elongated, and monophilic network topology}

For diverse oscillator dynamics and sufficient coupling budget, our gradient-based optimization robustly yields networks that support strong collective synchrony. Such optimal networks exhibit four structural patterns. They are \textbf{sparse}, with most edges eliminated, and \textbf{bipartite}, partitioning nodes by intrinsic frequency (e.g., each of the few remaining edges in Fig.~\ref{fig:network} connects a blue node to a red one). They are \textbf{elongated}, characterized by long paths that contrast the conventional assumption that short paths facilitate stronger synchrony~\cite{hong2002synchronization, zhao2006relations, brede2008synchrony, kelly2011topology}. During the course of optimization, the network structure stretches out, approaching a one-dimensional manifold topology \cite{taylor2015topological}.
Finally, they are extremely \textbf{monophilic} \cite{altenburger2018monophily, evtushenko2021paradox}, with neighbors of any node exhibiting highly similar frequencies that differ from the node’s own, reflecting specific preferences in neighbor selection (Fig.~\ref{fig:network}).

We find these four patterns in each of the five canonical heterogeneous oscillator models we examined, spanning a broad range of dynamical regimes (see the bottom row of Fig.~\ref{fig:network}): the Kuramoto model \cite{kuramoto1975international, strogatz2000kuramoto, arenas2008synchronization}, the seminal phase-oscillator system; the Sakaguchi--Kuramoto model \cite{sakaguchi1986soluble}, which introduces phase frustration, breaking the antisymmetry of the coupling; the swing equations \cite{filatrella2008analysis}, derived from electrically coupled generator turbines and incorporating inertia via a second-order time derivative; the Stuart--Landau model \cite{matthews1990phase}, describing coupled near-Hopf oscillators with two-dimensional phase--amplitude states; and coupled R\"ossler oscillators \cite{leyva2012explosive, skardal2017optimal}, which evolve on three-dimensional state spaces and exhibit deterministic chaos. Model definitions are given in Methods. Despite this diversity, spanning first- and second-order dynamics, phase-only and phase--amplitude coupling, and periodic and chaotic behavior, optimization consistently yields networks exhibiting all four patterns, with each emerging robustly during training. This consistency suggests these are general structural hallmarks of synchrony-optimal networks under a constrained coupling budget and uniform coupling costs.

The frequency-based bipartite coupling reflects an effective use of the constrained coupling budget by strategically balancing oscillators with opposing natural frequencies. This property helps explain prior  research that observed synchronization is enhanced by negative correlations between the frequencies of coupled nodes \cite{brede2008synchrony,buzna2009synchronization,skardal2014optimal,pinto2015optimal}. Such pairing amplifies edge effectiveness by ensuring the existence of a phase gap across connected nodes, thereby enhancing the contribution of each weighted connection. 

Monophily can be similarly explained by the optimal allocation of limited resources. The marginal benefit to synchrony $r$ per unit of edge weight depends sensitively on the frequency pairing of connected nodes. Hence, as optimization progresses, possible connections of a node with frequency $\omega$ tighten around the optimal neighbor having frequency $\nu(\omega)$ (which maximizes this marginal gain).

\edit{
Bipartition and monophily, together with continuously distributed frequencies, account for sparsity and elongation. Under strong monophily, each node couples only to a narrow band of frequencies, which for a continuous distribution contains only a small fraction of the nodes, so most possible connections are forgone. Elongation follows because the two endpoints of any two-hop path are both neighbors of the intermediate node, so they share a frequency sign and have similar magnitudes. Each two-hop step therefore moves only slightly across a continuum of frequencies, and getting from a small positive frequency to a large one takes many hops.}



\subsection*{Analytically tractable model}
To develop analytical insight into the emergence of these synchrony-optimizing structural patterns, we next restrict our attention to a tractable class of dynamics governed by the differential equation
\begin{equation}\label{oscillator_eq}
\frac{d\theta_i}{dt} = \omega_i - \sum_{j=1}^{N} A_{ij} H(\theta_i, \theta_j).
\end{equation}
Here, $\theta_i$ is the phase of node $i$ and $\omega_i$ is its intrinsic frequency, sampled from a mean-zero distribution $g(\omega)$. The network topology and edge weights are encoded by a nonnegative, symmetric adjacency matrix $\boldsymbol A$.
Interactions are mediated by an antisymmetric coupling function $H$. In systems with phase-difference coupling, $H(\theta_i, \theta_j) = H(\theta_i - \theta_j)$, rotation of the reference frame is a symmetry transformation that allows the mean frequency to be shifted arbitrarily ($\theta_i(t)\mapsto \theta_i(t)+\Omega t$), thereby relaxing the mean-zero assumption on $g(\omega)$. 

We specialize the subsequent numerical experiments and part of the analysis to the Kuramoto model, Eq.~\eqref{oscillator_eq}, with coupling function $H(\theta_i, \theta_j) = \sin(\theta_i - \theta_j)$. Phase synchronization is quantified by the Kuramoto order parameter
\begin{equation}\label{order_param}
    r(t) = \frac{1}{N} \left|\sum_{j=1}^{N} e^{i\theta_j(t)}\right|.
\end{equation}

\begin{figure*}[ht]
    \centering
    \includegraphics[width=\textwidth]{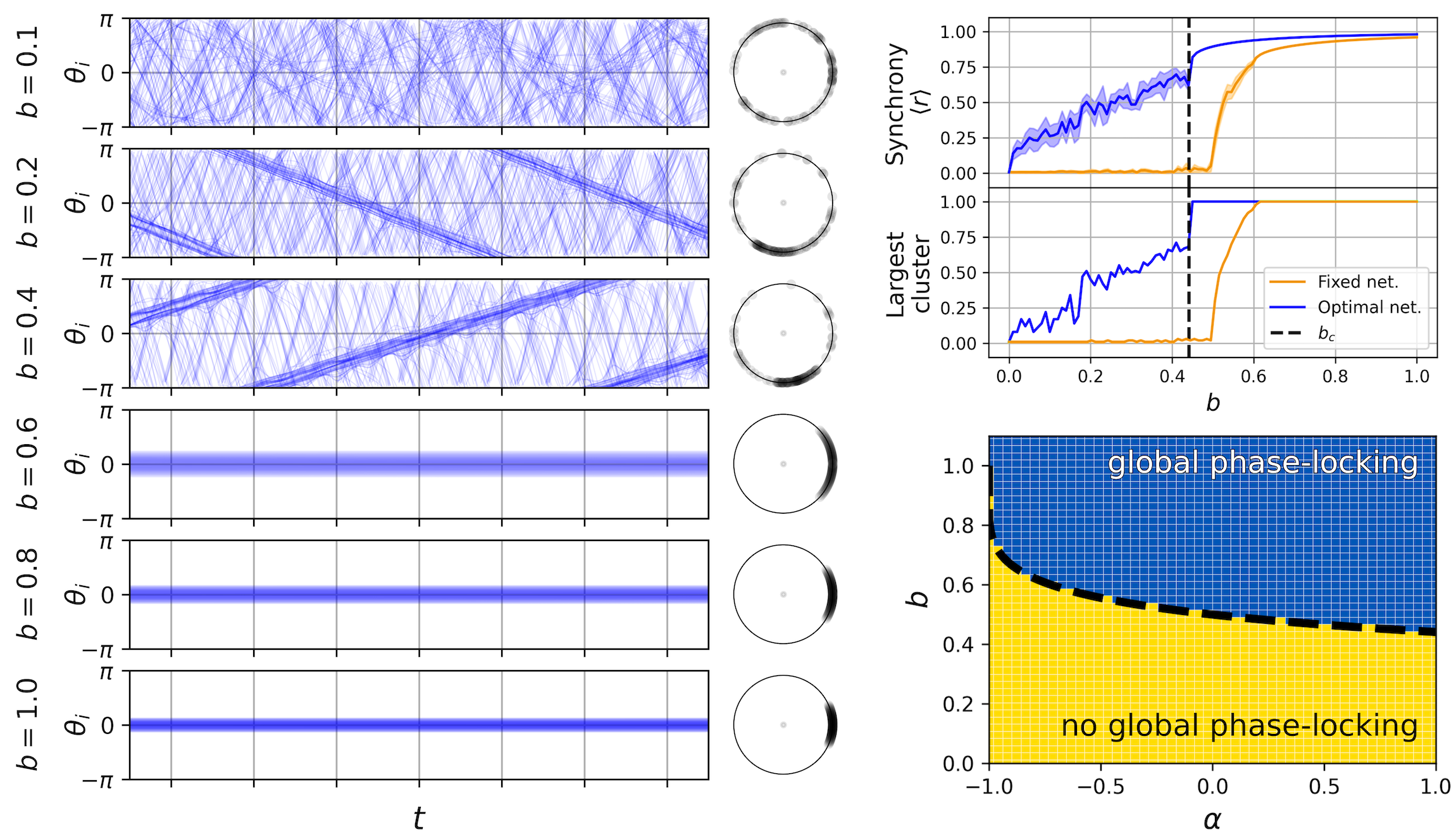}
    \caption{
        \textbf{Disappearance of the synchronization threshold and emergence of global phase-locking.}
        \textbf{(Left)} Time series of phases $\theta_i(t)$ for a synchrony-optimized Kuramoto network (and final phase snapshots on the unit circle) are shown for a varying coupling budget $b$. A cluster of phases exists for all $b$, and global phase-locking occurs above a critical budget $b_c$.
        \textbf{(Top-right)}~We plot the time-averaged Kuramoto order parameter and size of the largest phase-locked cluster versus $b$ for both a fixed all-to-all network (orange) and optimal networks (blue). Optimal networks lack a 
        synchronization threshold, and $b_c$ is shown by the   dashed black line given by Eq.~\eqref{b_c}.
        \textbf{(Bottom-right)} Phase diagram showing the absence/presence of global phase-locking for optimal networks as a function of  coupling budget $b$ and a frequency distribution parameter $\alpha$.
        Color indicates the outcome of numerical optimization and simulation.
        In all panels, $N=100$ intrinsic frequencies $\omega \in [-1, 1]$ are drawn from the family $g_\alpha(\omega) \propto 1/(1 + \alpha \omega^2)$. For this family, Eq.~\eqref{b_c} yields $b_c = \log(1+\alpha) / (2\sqrt{\alpha} \arctan\sqrt{\alpha})$. The bottom-right diagram varies $\alpha \in [-1, 1]$, sweeping through bimodal and unimodal frequency distributions, while the remaining panels fix $\alpha = 1$.
    }
    \label{fig:phase_diagrams}
\end{figure*}

\subsection*{Dynamical regimes of optimal networks}
Before deriving optimal network structure analytically, we first describe how the global dynamical behavior depends on the coupling budget $b$. As summarized in Fig.~\ref{fig:phase_diagrams}, two regimes emerge: persistent partial synchrony at small $b$, and global phase-locking above a critical budget $b_c$. The four structural patterns and the analytical results that follow both pertain to this supercritical regime.

The dynamical behavior of optimized systems with increasing budget departs markedly from the classical case of a fixed network with varying coupling strength. The conventional threshold of synchronization is eliminated, giving way to a regime of persistent nonzero synchrony even under arbitrarily weak coupling. This phenomenon is demonstrated numerically in the top-right panel of Fig.~\ref{fig:phase_diagrams} and proved analytically in the Methods. The mechanism is structural: rather than spreading the budget uniformly, optimization concentrates it on a strategically chosen subset of edges, and it is this concentration that sustains synchrony in the weak-coupling regime.

We note that prior research has identified one mechanism that can cause the synchronization threshold to disappear: asymptotically large random networks in which the node degrees diverge with network size \cite{restrepo2005onset}. Here, we show that, for optimal networks, the synchronization threshold also disappears without requiring node strengths (the weighted-network analog of node degrees) to become infinitely large.

A distinct dynamical phase transition emerges: at a critical budget, the system attains global phase-locking, with all oscillators asymptotically converging to a static stationary state (see Fig.~\ref{fig:phase_diagrams}).

The critical budget can be characterized by a lower bound on the edge weights required to support a static solution of Eq.~\eqref{oscillator_eq}, which optimization saturates (see Methods). In the thermodynamic limit $N \to \infty$, this yields a critical budget
\begin{equation}\label{b_c}
b_c = \frac{2}{\max(H)} \int_0^{\infty} \omega g(\omega) d\omega,
\end{equation}
which is an expression that depends explicitly on the coupling function $H$ and   frequency distribution $g(\omega)$. The difficulty of phase-locking increases with the mean of the positive frequencies. Owing to the mean-zero condition on $g(\omega)$, this is equivalently reflected in the magnitude of the negative-frequency contribution. Equation~\eqref{b_c} is validated through numerical optimization experiments (see black dashed line in Fig.~\ref{fig:phase_diagrams}).

\begin{figure*}[!t]
    \centering
    \includegraphics[width=\textwidth]{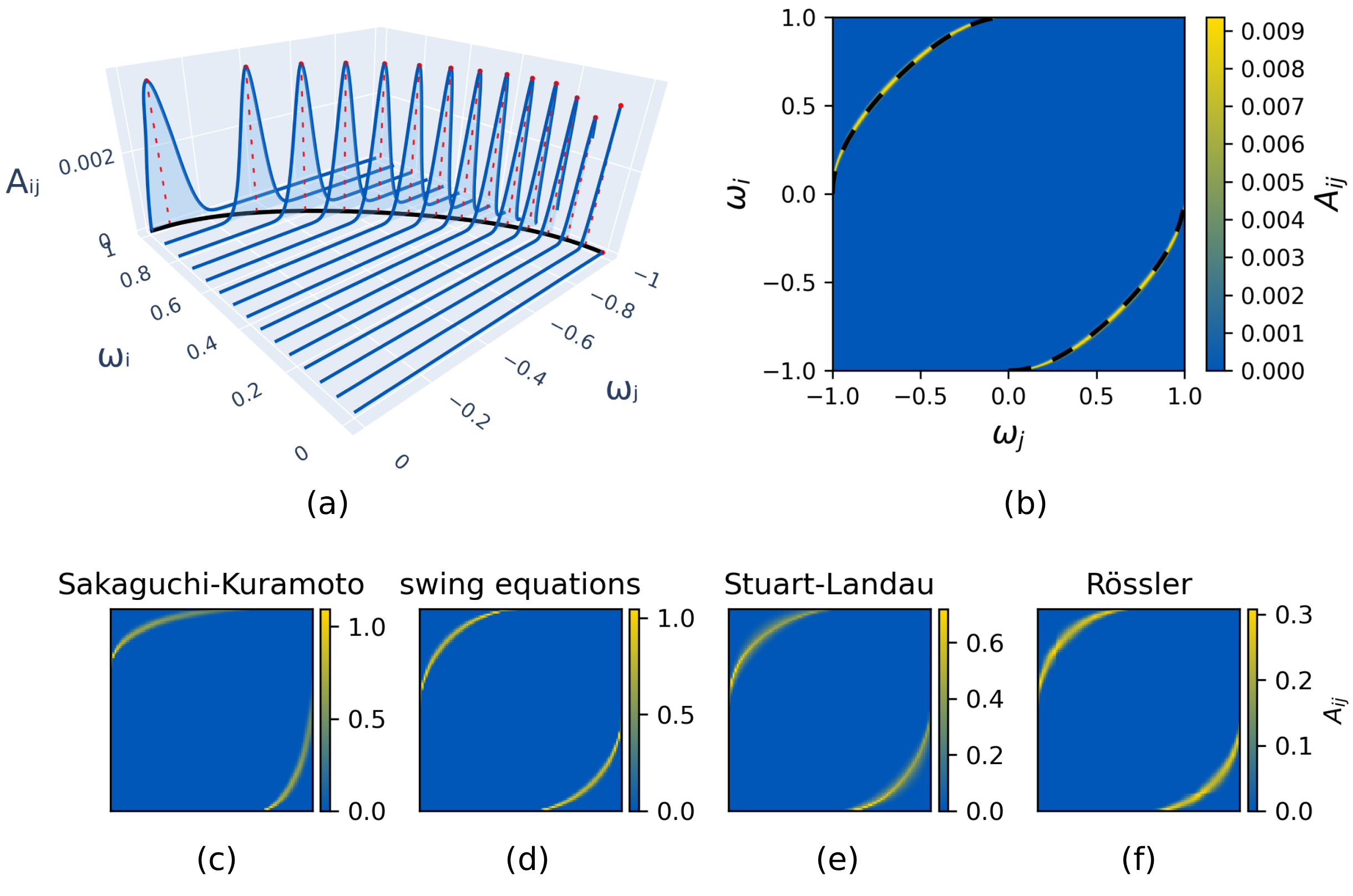}
    \caption{ 
        \textbf{Edge weights $A_{ij}$ as a function of oscillator frequencies} $\omega_i$ and $\omega_j$ in a synchrony-optimized Kuramoto network. Black curves in either subplot mark the negative solution  $\omega_j = \nu_-(\omega_i)$ of Eq.~\eqref{curve_eq}.
        \textbf{(a)} Each blue curve traces the distribution of neighbor frequencies for a given $\omega_i$, with density peaking near the predicted $\nu_-$. Results are shown after 2000 optimization steps.
        \textbf{(b)} A heatmap representation of the network after 5000 additional steps, where the alignment sharpens significantly within the frequency$\times$frequency space and closely matches our analytical prediction $\nu_-(\omega)$ (dashed black curves).
        Both panels use   $N=10^4$,   $b=1$, and   $g_1(\omega) \propto 1/(1 + \omega^2)$.
        Bottom row: optimized adjacency matrices for oscillators with uniformly distributed frequencies in \textbf{(c)} Sakaguchi–Kuramoto model, \textbf{(d)} power grid swing equations, \textbf{(e)} Stuart–Landau oscillators, and \textbf{(f)} chaotic R\"ossler oscillators. See Methods section for definitions.
    }\label{fig:adjacencies}
\end{figure*}

\begin{figure*}[t!]
    \centering
    \includegraphics[width=\textwidth]{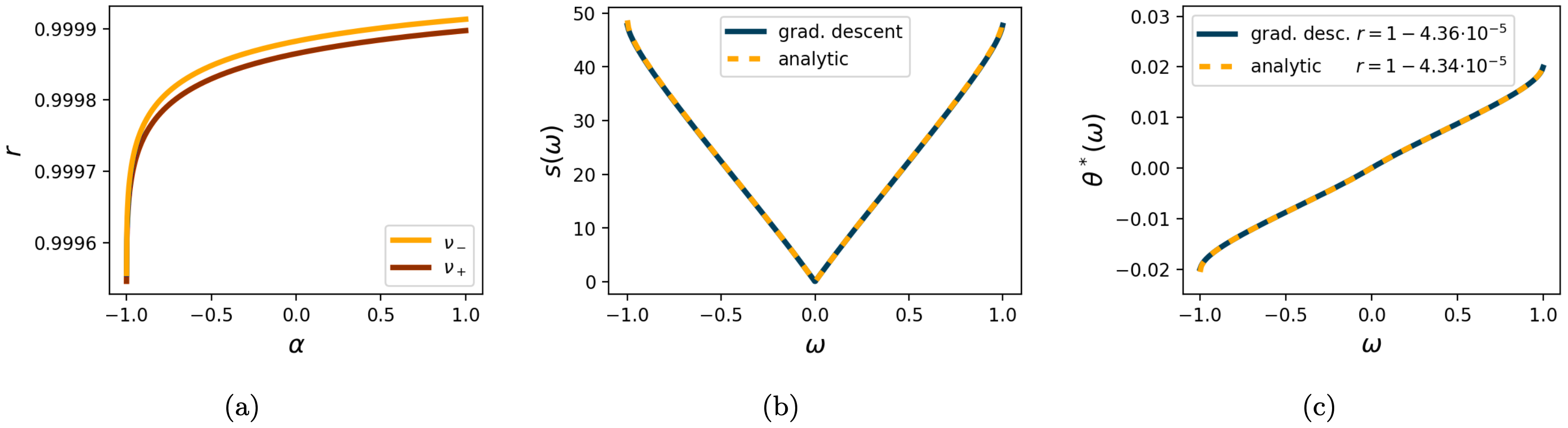}
    \caption{
        \textbf{Analytical and numerical results for the strong-coupling regime.}
        \textbf{(a)} Kuramoto order parameter $r$ from Eq.\eqref{eq:strongly_coupled:r} evaluated using the two pairing function branches $\nu_+(\omega)$ and $\nu_-(\omega)$ across the family of frequency distributions  $g_\alpha(\omega) \propto 1/(1 + \alpha \omega^2)$. The nontrivial branch $\nu_-$ consistently outperforms $\nu_+$, explaining the convergence of our gradient-based optimization to $\nu_-$.
        \textbf{(b)}~Node strengths $s(\omega)$ from Eq.\eqref{eq:strongly_coupled:s} (dashed) compared with optimized results (solid) after $10^5$ training steps with $N=10^4$, $b=20$, and $\alpha=1$.
        \textbf{(c)} Stationary phases $\theta^*(\omega)$ from Eq.~\eqref{eq:strongly_coupled:theta} (dashed) versus simulation on the numerically optimized network (solid). These reported $r$ values indicate near-perfect agreement between analytical and numerical solutions, with the analytical prediction being slightly higher than the numerics.
    }
    \label{fig:strong_coupling_results}
\end{figure*}

\subsection*{Optimal pairing function $\nu(\omega)$}

The four structural patterns described above prove to be amenable to detailed analytical characterization. Consider a node with intrinsic frequency $\omega$, and let $\nu(\omega)$ denote the mean frequency of its neighbors. The consistently observed strong monophily means that the neighbors' frequency distribution has a narrow standard deviation: $\omega$ effectively only connects to nodes with frequencies very close to $\nu(\omega)$.
We propose a general ansatz for the adjacency matrix that captures this property. By invoking symmetry and antisymmetry constraints, and applying algebraic and asymptotic analysis (see Methods), we derive a compact condition for the form of the pairing function:
\begin{equation} \label{bipartite}
    \omega \nu \le 0,
\end{equation}
\begin{equation} \label{curve_eq}
    \omega g(\omega) d\omega = \pm \nu g(\nu)d\nu.
\end{equation}
The first condition, Eq.~\eqref{bipartite}, enforces sign-segregated bipartite connectivity, a structural feature long recognized as beneficial, but whose importance we establish here analytically. 

The second condition, Eq.~\eqref{curve_eq},  is a nonlinear differential equation that enforces a self-consistent stationary balance between each oscillator’s intrinsic frequency and the coupling terms from its neighbors, such that mutual interactions are exactly counterbalanced.
Equation~\eqref{curve_eq} enforces this balance for an infinitesimal band of frequencies $d\omega$ around $\omega$, and their paired counterparts $d\nu(\omega)$ around $\nu(\omega)$. Each side of Eq.~\eqref{curve_eq} reflects the total coupling influence that the network must supply to hold that infinitesimal band in equilibrium, where $g(\omega)d\omega$ describes the number of such oscillators, and $\omega$ gives the tendency of each one to drift away from the locked state. The sign term comes from the fact that $\nu(\omega)$ could be either an increasing or a decreasing function.

Equation \eqref{curve_eq} admits two solutions corresponding to the $\pm$ sign, which we denote $\nu_+$ and $\nu_-$ respectively. The positive branch, $\nu_+$, pairs high-frequency oscillators with others of similar magnitude---fast positives with fast negatives, slow positives with slow negatives. In contrast, the negative branch, $\nu_-$, links the fast negative-frequency nodes with the slow positive-frequency nodes, and vice versa. For symmetric frequency distributions, $g(-\omega) = g(\omega)$, the positive solution is trivial $\nu_+(\omega) = -\omega$, providing a theoretical derivation to prior simulated observations \cite{kelly2011topology}. In contrast, other papers report characteristics similar to that of the negative branch (See, e.g., Fig.~4(d) in \cite{skardal2014optimal}, Fig.~3 in \cite{pinto2015optimal}, Fig.~4(k) in \cite{papadopoulos2017development}, and Fig.~3(g) in \cite{brede2018competitive}.) Our gradient-based optimization reconciles these two contrasting solutions and resolves the literature conflict by consistently converging to the negative branch $\nu_-(\omega)$ (see Fig.~\ref{fig:adjacencies}). We will later show analytically that $\nu_+(\omega)$ is suboptimal compared to $\nu_-(\omega)$ for the strongly coupled Kuramoto model across a diverse family of frequency distributions. The core advantage of $\nu_-$ is that it pairs nodes with significantly different frequencies, ensuring connections always bridge meaningful phase differences. In contrast, $\nu_+$ wastes resources by connecting oscillators with small positive frequencies with those having  small negative  frequencies despite their phase differences being small.

\subsection*{Optimal strength allocation $s(\omega)$}

While the pairing function $\nu(\omega)$ specifies the network topology by identifying which node pairs are connected,
it leaves unspecified how the coupling budget is allocated across those links. To characterize how the coupling budget is optimally distributed, we analyze the node strength $s(\omega)$, defined as the total edge weight incident to a node with intrinsic frequency $\omega$.

This problem admits a natural formulation as a constrained variational principle. Specializing to Kuramoto dynamics and optimizing $r$ in the globally phase-locked stationary state, we construct a Lagrangian encoding both the order parameter and the constraint on the coupling budget (see Methods). Assuming interactions are narrowly concentrated along the pairing function $\nu(\omega)$, we derive an Euler–Lagrange equation for $s(\omega)$:
\begin{equation}\label{s_condition}
    \frac{\omega^6}{s^6}
    =
    c \left(1-\frac{\omega^2}{s^2} \right)
    \left(\nu^2+\omega^2-2\omega\nu\sqrt{1-\frac{\omega^2}{s^2}} \right),
\end{equation}
where $c$ is a Lagrange multiplier enforcing the coupling budget constraint. While solving Eq.~\eqref{s_condition} generally requires numerical methods, we analyze $s(\omega)$ analytically just above the critical point $b_c$ and in the strongly coupled regime. For more realistic dynamical systems, we expect the variational equations to become substantially more complex, or even intractable. We now analyze Eq.~\eqref{s_condition} in two tractable limits.

\begin{figure*}[t!]
    \centering
    \includegraphics[width=\textwidth]{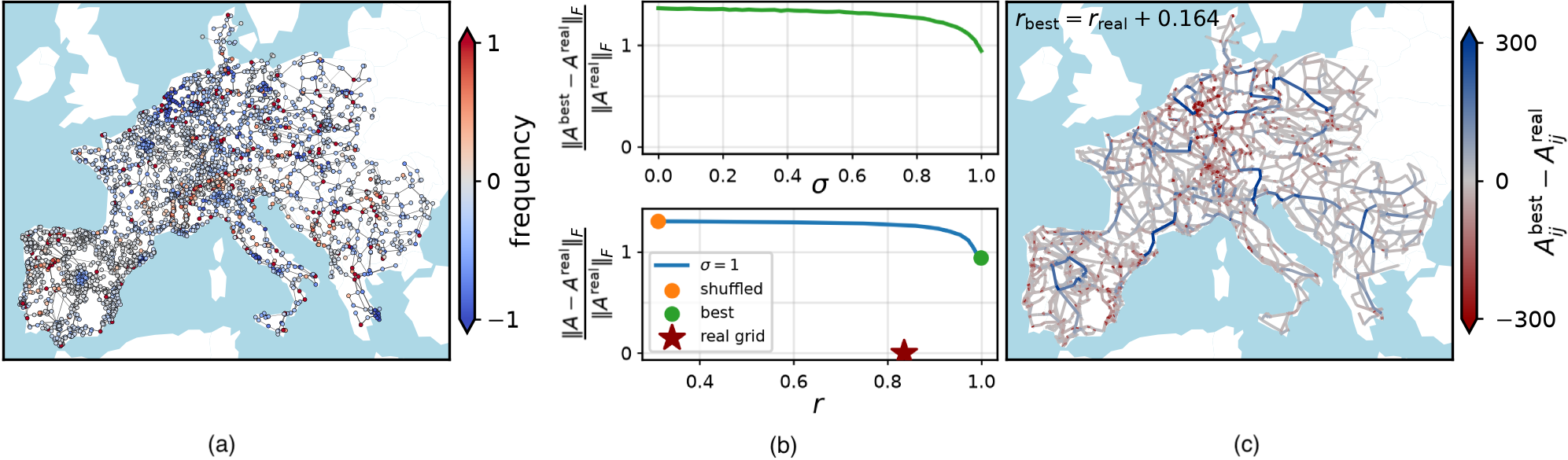}
    \caption{\edit{
        \textbf{Optimizing the continental European power grid for synchrony.}
        \textbf{(a)} PanTaGruEl reconstruction of the synchronous continental European grid \cite{pagnier2019inertia,pagnier_2022_5816149}. Nodes are colored by intrinsic frequency $\omega_i$, obtained by centering the net active-power injections and normalizing them to unit variance. \textbf{(b)} Comparison of the empirical and synchrony-optimized edge weights. 
        The sparse topology is held fixed, and optimization preserves the empirical cost budget $b_\sigma=N^{-1}\sum_{ij}c_\sigma(i,j)A_{ij}$, with $c_\sigma(i,j)=(1-\sigma)+\sigma d_{ij}$, and $d_{ij}$ is the normalized great-circle distance between the endpoints of each line.
        The parameter $\sigma$ interpolates between uniform costs ($\sigma=0$) and length-proportional costs ($\sigma=1$).
        \emph{Top:} Relative distance $\|\mathbf A^{\mathrm{best}}-\mathbf A^{\mathrm{real}}\|_F/  \|\mathbf A^{\mathrm{real}}\|_F$ between the optimized and empirical edge weights versus  $\sigma$. 
        \emph{Bottom:} Optimization trajectory for $\sigma=1$, shown in terms of relative distance from the empirical weights and synchronization $r$. Optimization begins from a random reallocation of the cost budget over  existing edges (orange circle); the empirical grid is marked by the red star and the optimized grid by the green circle. 
        \textbf{(c)} Edge-weight difference $A^{\mathrm{best}}_{ij}-A^{\mathrm{real}}_{ij}$ for $\sigma=1$, for which $r_{\mathrm{best}}-r_{\mathrm{real}}=0.164$. Observe  edge-weight increases
        %
        occur predominantly for longer power  lines (blue), whereas decreases occur for shorter ones (red).
        }
        }
    \label{fig:pantagruel}
\end{figure*}

\subsection*{System behavior at the critical point}
A necessary condition for Eq.~\eqref{s_condition} to admit a real solution is $s(\omega) \geq |\omega|$. To analyze the critical behavior of Eq.~\eqref{s_condition}, we perform a leading-order Taylor expansion of $s(\omega)$ about $|\omega|$  (see Methods) to obtain

\begin{align}
    r &= r_c+\kappa(b-b_c)^{1/2},
    \label{eq:weak:r} 
    \\
    r_c &= \int_0^\infty (\omega^{-2}+\nu^{-2})^{1/2} \omega g(\omega) d\omega,
    \label{eq:weak:rc}
    \\
    b_c &= 2\int_0^\infty \omega g(\omega)d\omega.
    \label{eq:weak:bc}
\end{align}
Here the positive constant coefficient is given by $\kappa^2=\tfrac{1}{2}\int_0^\infty (\omega^2+\nu^2)^{-1}\omega g(\omega)d\omega$. 
Equations~\eqref{eq:weak:r}-\eqref{eq:weak:bc}
characterize the phase-locking transition: For $b<b_c$ the system is not fully phase-locked. When $b\ge b_c$, the system has enough coupling budget to ensure full phase-locking, which induces a sharp surge in synchrony $r$ just after the transition.
Equation~\eqref{eq:weak:r} describes the critical scaling of $r$ just above the transition, exhibiting a power-law behavior with exponent $1/2$. Equation~\eqref{eq:weak:rc} gives the critical synchrony level $r_c$ at the onset of global phase-locking, which is used in Methods to prove the absence of a synchronization threshold. The derivation also yields Eq.~\eqref{eq:weak:bc}, which recovers Eq.~\eqref{b_c} for the Kuramoto coupling and elucidates its emergence from first principles.

\subsection*{Solution in the strongly coupled regime}
When the coupling budget is large, node strengths significantly exceed corresponding frequency magnitudes, i.e., $s(\omega) \gg |\omega|$. In this regime, Eq.~(\ref{s_condition}) simplifies, alongside the stationary state and order parameter equations, yielding a closed-form leading-order solution (see Methods):
\begin{align}
    r &= 1 - \frac{\chi^{3}}{4b^{2}}, \label{eq:strongly_coupled:r} \\
    s(\omega) &= \frac{b}{\chi} \frac{|\omega|}{|\omega - \nu|^{1/3}}, \label{eq:strongly_coupled:s} \\
    \theta^*(\omega) &= \frac{\chi}{b} \frac{\omega}{|\omega - \nu|^{2/3}}, \label{eq:strongly_coupled:theta}
\end{align}
where we defined $\chi\coloneq 2\int_{0}^{\infty}(\omega-\nu)^{-1/3}\omega g(\omega)d\omega$.

Equation~\eqref{eq:strongly_coupled:r} reveals a power-law approach to perfect synchrony, with $1 - r \propto b^{-2}$. It also enables a direct comparison between the synchrony-enhancing efficacy of the two pairing function branches, $\nu_+(\omega)$ and $\nu_-(\omega)$, as the derivation applies to either.
Across the family of frequency distributions $g_\alpha(\omega) \propto 1/(1 + \alpha \omega^2)$, the nontrivial branch $\nu_-(\omega)$ consistently yields higher synchrony (Fig.~\ref{fig:strong_coupling_results}(a)), explaining the exclusive convergence of gradient-based optimization to this branch.

Equation~\eqref{eq:strongly_coupled:s} and the corresponding plot in Fig.~\ref{fig:strong_coupling_results}(b) reveal that node strength increases with the magnitude of intrinsic frequency, indicating that faster oscillators receive proportionally more coupling resources—a correlation long recognized heuristically, but for which we provide the first analytical explanation. Similarly, Eq.\eqref{eq:strongly_coupled:theta} and Fig.~\ref{fig:strong_coupling_results}(c) show that stationary phases increase monotonically with frequency, consistent with the expectation that faster oscillators lead in phase relative to slower ones.

Gradient computations also simplify in the strongly coupled regime. The gradient of the linearized order parameter, i.e., the synchrony alignment function (SAF) \cite{skardal2014optimal} admits a closed-form solution \cite{taylor2016synchronization}. We leverage this analytical gradient to accelerate optimization by around one order of magnitude. The optimized structural and phase configurations closely match the analytic solution derived in Eqs.~\eqref{eq:strongly_coupled:r}-\eqref{eq:strongly_coupled:theta}, as shown in Figs.~\ref{fig:strong_coupling_results}(b) and \ref{fig:strong_coupling_results}(c).

\edit{
\subsection*{Optimizing a spatial power grid for synchrony}

Synchrony is only one of many objectives shaping real power grids, which must also deliver energy to spatially dispersed consumers and withstand shocks and cascading outages, all under spatially heterogeneous costs, transmission losses, existing rights-of-way, and economic and political constraints that span national borders. We therefore do not expect an empirical grid to exhibit the structural patterns described above. Instead, we ask how close a real grid is to the synchrony optimum once its spatial layout is taken into account, and how its existing coupling would be reallocated if synchrony were the sole objective.

As discussed in the Methods, we consider the continental European grid from the PanTaGruEl dataset \cite{pagnier2019inertia,pagnier_2022_5816149} (Fig.~\ref{fig:pantagruel}(a)). Keeping the observed topology fixed, we optimize the weights of the existing power lines under a fixed budget in which the cost of an edge depends on its length. We introduce a parameter $\sigma$ that interpolates between uniform costs ($\sigma=0$) and length-proportional costs ($\sigma=1$).

As length-proportional costs gain weight with increasing $\sigma$, the synchrony-optimal edge-weight allocation approaches the empirical one (Fig.~\ref{fig:pantagruel}(b), top), indicating that length-proportional costs describe this grid better than uniform costs. The empirical allocation is also substantially more synchronizable than a random reallocation of the same cost budget, and optimization starting from such a random allocation moves the network toward the real grid (Fig.~\ref{fig:pantagruel}(b), bottom). Together, these observations suggest that, once its spatial layout is taken into account, the continental European grid optimizes synchrony to an extent.

The real grid nevertheless remains distinct from the synchrony optimum. At equal total cost, optimization increases $r$ by $0.164$. This improvement comes primarily from shifting coupling away from shorter, local lines and toward longer-range lines (Fig.~\ref{fig:pantagruel}(c)), even though the latter are more expensive under length-proportional costs. Within this cost model, long-distance transmission therefore appears under-invested relative to its value for synchrony, consistent with findings that long-range interlinks improve synchronization stability in the eastern Japan grid~\cite{wang2016enhancing}.
}

\section*{Discussion}

While the dynamics and phase transitions of lattices and random graphs have been extensively studied, those of optimal networks remain  underexplored.
A main challenge is that an optimal network is itself often analytically and computationally intractable to obtain.
Here, we have shown that such studies are now much more tractable with the advent of modern computational tools for machine learning. Hardware and software designed to optimize neural networks can be repurposed to optimize other network processes, opening up new research avenues to pursue ``optimal network science.'' 

Our open-source repository \cite{mikaberidze2025networkoptimization}, further developed into a customizable Python package \cite{mikaberidze2026gradnet}, implements a scalable, differentiable optimization framework in \textsc{PyTorch} and \textsc{torchdiffeq}, enforcing coupling budget constraints and supporting gradient-based tuning of network structure. This framework, when applied to five canonical models of synchrony, uncovered that a hallmark of synchrony-optimal structures is extreme monophily: the neighboring frequencies for any oscillator with frequency $\omega$ are tightly concentrated around some frequency $\nu(\omega)$. \edit{Together with bipartition, monophily is sufficient to account for the two remaining patterns: sparsity follows because each node's admissible neighbors are confined to a narrow frequency band, and elongation follows because two-hop paths remain trapped among similarly-signed, similarly-sized frequencies, forcing long detours across the frequency spectrum.} These structural patterns were found to persist across oscillator models spanning phase-only dynamics with and without frustration, inertial systems, phase–amplitude coupling, and deterministic chaos.

Motivated by this observation, we derived a ``constructive'' characterization of synchrony-optimal networks for oscillatory dynamical systems given by Eq.~\eqref{oscillator_eq}. A self-consistent differential equation [Eq.~\eqref{curve_eq}] reveals which pairs of oscillators are to be connected, while an Euler–Lagrange equation [Eq.~\eqref{s_condition}], tailored specifically to the Kuramoto model, indicates how much edge weight is assigned to each oscillator through its node strength $s(\omega)$. Note that the derivations are performed in the thermodynamic limit. Nevertheless, our numerical experiments indicate that the key structural patterns persist and remain quantitatively accurate even for very small system sizes (e.g., $N=20$).
This analysis elevates prior numerical observations about structures that enhance synchronization (e.g., bipartition and  frequency–strength correlations), and  it rules out other observations (e.g., small diameters and homogeneity) \cite{hong2002synchronization, nishikawa2003heterogeneity, zhao2006relations, brede2008synchrony, grabow2010small, kelly2011topology}. It also reconciles conflicting reports in the literature \cite{donetti2005entangled, brede2008synchrony, brede2018competitive,skardal2014optimal, taylor2016synchronization, chamlagai2022grass} regarding autocorrelation of frequency magnitudes, unifying them under a single theoretical framework with two solutions: $\nu_-$ and $\nu_+$.

This constructive characterization also enabled analysis of the synchronization dynamics unfolding over such structures, including closed-form expressions for the critical point $b_c$ of global phase-locking [Eq.~\eqref{b_c}], the scaling laws just above $b_c$ [Eqs.~\eqref{eq:weak:r}--\eqref{eq:weak:bc}], and limiting behavior in the strong-synchrony regime [Eqs.~\eqref{eq:strongly_coupled:r}--\eqref{eq:strongly_coupled:theta}, Fig.~\ref{fig:strong_coupling_results}]. Additionally, we found that optimal networks lack a synchronization threshold (see Methods).

%
%
%


Our proposed computational and analytical techniques, as well as our discovery of the four structural patterns may have implications for guiding the engineering of real-world systems in which synchronization is essential. For power-grid dynamics governed by swing equations \cite{filatrella2008analysis}, modifying a single transmission line can either enhance or impair global synchrony \cite{witthaut2012braess, motter2013spontaneous, wang2016enhancing, schafer2022understanding}, underscoring the importance of strategic resource allocation.
\edit{Indeed, our preliminary analysis of the continental European grid partially explains its structure by synchrony optimization, and indicates that synchrony can be further improved by shifting investment from local lines to specific long-range connections.} 
The Stuart–Landau model \cite{matthews1990phase}, a normal form underpinning a wide class of physical, chemical, and biological oscillators, suggests potential applications including synchronizing laser arrays \cite{nixon2012controlling,ye2025optimal}, neuromorphic computing platforms based on coupled oscillators \cite{romera2018vowel, romera2022binding, torrejon2017neuromorphic}, and coherently coupled sensor arrays \cite{antonio2012frequency}, all of which rely on synchrony for efficient functioning. Because these structural patterns have not previously been articulated, existing engineered platforms are unlikely to already embody them. Our work thus offers a forward-looking design blueprint, with direct experimental tests requiring purpose-built networks. Coupled Rössler oscillators offer one such platform, with straightforward circuit implementations and programmable couplings \cite{leyva2012explosive,skardal2017optimal}, providing an accessible route to experimental verification. 

It is also interesting to consider whether these structural hallmarks  arise in naturally evolved systems. 
For example, Eq.~\eqref{oscillator_eq} encompasses models of consensus-based decision making in social networks \cite{olfati2007consensus, kia2019tutorial, kamthe2025external}. Monophily, a pattern widely observed in social networks \cite{altenburger2018monophily, evtushenko2021paradox}, is typically explained mechanistically, by appeal to ``individuals with extreme preferences for a particular attribute possibly unrelated to their own attribute'' \cite{altenburger2018monophily}. Our findings suggest a complementary, functional explanation: monophily promotes efficient consensus.
Another potentially fruitful direction is the study of heterogeneous neurons in the suprachiasmatic nucleus, where synchronization is crucial for establishing circadian rhythms \cite{yamaguchi2003synchronization, nikhil2025inferred}. However, it should be emphasized that naturally occurring networks have been subjected to complicated, multi-objective evolutionary pressures, whereas here we have focused on optimizing a single order parameter $r$. To help extend our findings to investigate naturally evolved networks, future work should consider the optimization of multi-objective settings where synchrony competes alongside robustness and other inherent real-world constraints \cite{onnela2011geographic, bullmore2012economy}.


We close by highlighting several other directions for extension. 
Building on our analysis of the supercritical regime, future studies could explore the subcritical domain characterized by cluster phase-locking \cite{daido1990intrinsic}. While the four structural patterns were found to be empirically robust across several important systems differing from  Eq.~\eqref{oscillator_eq}, a theoretical understanding of why they persist, when they break, and whether a pairing function or strength distributions can be derived for them, remains open. Our framework also enables systematic exploration across diverse antisymmetric coupling forms (e.g., second and higher-harmonic couplings  \cite{skardal2011cluster, komarov2013multiplicity}), and their exploration would test the theory’s full generality. It would also be interesting to explore the optimization of measures of synchrony beyond the   order parameter $r$. Extended work could optimize the convergence rate of relaxation dynamics \cite{timme2006speed, witthaut2022collective}, decentralized and multi-level notions of synchrony \cite{rohden2014impact, chamlagai2022grass}, the size of basins of attraction \cite{wiley2006size, menck2013basin}, and the properties and controllability of induced chimera states \cite{panaggio2015chimera, omelchenko2016tweezers}. \edit{Further study should also address synchrony optimization in spatially embedded networks, where spatial layout may favor or hinder synchrony and structural patterns such as monophily must be generalized to account jointly for node position and intrinsic frequency.} 
The algorithmic and analytical tools presented herein offer fruitful directions to make progress on these fronts and may generalize to other classes of dynamical processes that have distinct types of phase transitions and collective objectives.



\section*{Methods}

\subsection*{Gradient-based network optimization}

Here, we present implementation details for a gradient-based optimization framework that takes advantage of modern computational tools for the optimization of neural networks for machine learning.
%
Our algorithm involves iterating three steps:
\begin{itemize}
    \item a ``forward pass'' in which we consider fixed  parameters $P_{ij}$  and numerically integrate Eq.~\eqref{oscillator_eq} and compute the time-average order parameter $\langle r\rangle$; 
    \item a ``backward pass'' (i.e., backpropagation) in which we compute the gradient  of $\langle r\rangle$ with respect to $P_{ij}$ using a chain rule with partial derivatives for intermediate variables (e.g., $A_{ij}$, ${\bf \theta}_t$, and $r_t$); and
    \item the parameters are updated, $ P_{ij} \mapsto P_{ij} + \eta \frac{d\langle r\rangle}{dP_{ij}}$, where learning rate $\eta$ controls the step size in the gradient direction.
\end{itemize}
The terminology forward/backward refers to these steps' associated ``computational graphs'' (Fig.~\ref{fig:computational_graph}), and each iteration is referred to as an ``epoch.''

Starting from a random (or heuristic) initialization of $P_{ij}$, we   iterate the above steps with the following specifications.

The forward pass begins by computing the adjacency matrix $\mathbf A$ from $\mathbf{P}$ using Eq.~\eqref{adj_from_P}. Then we integrate Eq.~\eqref{oscillator_eq} using \textsc{PyTorch}-based numerical integrator \textsc{torchdiffeq} \cite{paszke2017automatic,torchdiffeq}. Integration begins at $\theta_i(0) = 0$, yielding a strong and stable optimization signal early in the optimization process, substantially improving convergence stability. At later stages of training, one can introduce a relaxation period during which the gradients are not tracked. We ensure reproducibility by deterministically sampling the intrinsic frequencies, ${\omega_i}$, from the distribution $g(\omega)$ using the inverse cumulative distribution function evaluated at evenly spaced quantiles.

During the backward pass, as illustrated in Fig.~\ref{fig:computational_graph}, we compute 
$\frac{d}{d P_{ij}}(\langle r\rangle)$ for each $i,j$ using a partial derivative expansion that is efficiently implemented by considering a ``computational graph'' of dependent variables (each of which is defined by a differentiable relation). That is,
the map from ${\bf P}$ to the weighted adjacency matrix ${\bf A}$ (i.e., Eq.~\eqref{adj_from_P}) is differentiable. Somewhat counterintuitively, each step of the numerical integration process is   differentiable \cite{chen2018neuralode}, as is the calculation for $\langle r\rangle$. 
We compute the gradients automatically by leveraging modern machine-learning libraries like \textsc{PyTorch} and \textsc{torchdiffeq}, eliminating the need for explicit application of the lengthy chain rule.

The parameters are then updated by following the gradient uphill. Although this step appears straightforward, it can be refined substantially. We employ the Adam optimizer, which adapts step sizes along different parameter directions and smooths updates, improving convergence and reducing sensitivity to local irregularities in the landscape. Additionally, we exploit a scale-invariance in the mapping from $\boldsymbol P$ to $\boldsymbol A$, observing that while $\boldsymbol A$ remains unchanged under global rescaling of $\boldsymbol P$, the corresponding gradients do not. Normalizing $\boldsymbol P$ to fixed Frobenius norm $||\boldsymbol P||_F = \sqrt{N}$ after each update step markedly improves training stability and efficiency. We use the Adam optimizer with a log-spaced learning rate schedule spanning from $10^{-1}$ to $10^{-4}$ and train for typically $1$ to $5$ thousand iterations. The dynamics are integrated over a time horizon of $150$ using a Dormand--Prince RK5(4) scheme with relative and absolute tolerances of ${\sim}\,10^{-4}$.

Before continuing, we note that the parameters $P_{ij}$ need not be initialized randomly. In low-budget regimes, numerical optimization can be sensitive to initialization; naive choices, such as uniform or random starting points, often hinder convergence or trap the solver in poor local optima. To address this, we leverage our analytical finding that the shape of the optimal pairing function $\nu(\omega)$ is independent of the budget $b$. This allows us to initialize $P_{ij}$ using the optimized solution from a higher-budget system, providing a well-informed starting point that, empirically, significantly stabilizes convergence and improves the likelihood of reaching the global optimum. That is, in this scenario analytical insights inform and improve our numerical optimization. Initializing optimization experiments with networks having bipartition or their predicted node strengths can similarly guide the numerics to further enhance convergence stability and speed.

\subsection*{Analytical gradients for strong coupling}

Significant numerical acceleration can be obtained in the strongly coupled Kuramoto model with very large budgets. To this end, one can assume a phase-locked state and linearize $r$ near perfect synchrony (i.e., $r\approx 1$) to  obtain the Synchrony Alignment Function (SAF)  \cite{skardal2014optimal} and its associated approximation
\begin{equation}\label{saf}
    r \approx 1 - \frac{1}{2N} 
    \sum_{j=2}^N \lambda_j^{-2} \langle\boldsymbol{v}^j, \boldsymbol\omega\rangle^2.
\end{equation}
Here, $\lambda_j$ and $\boldsymbol v^j$ are the j-th eigenvalue and eigenvector of the Laplacian matrix ${\bf L} = {\bf D}-{\bf A}$ with $D_{ii}=\sum_j A_{ij}$. Vector $\boldsymbol \omega$   encodes the oscillator frequencies $\omega_i$, and $\langle\cdot,\cdot\rangle$ denotes the inner product. 

As shown in \cite{taylor2016synchronization},  Eq.~\eqref{saf} admits a closed-form expression for the gradient with respect to the edge weights
\begin{equation}\label{nabla}
    \nabla_{pq}r=-\frac{1}{N} 
    \sum_{j=2}^N 
        \frac{
            (v^j_p-v^j_q)
            \langle\boldsymbol{v}^j, \boldsymbol\omega\rangle
        }{
            \lambda_j^3
        }
    \sum_{i=1}^N 
        \frac{
            (v^i_p-v^i_q)
            \langle\boldsymbol{v}^i, \boldsymbol\omega\rangle
        }{
            1-\lambda_i/\lambda_j - \delta_{ij}
        }.
\end{equation}
Here $\nabla_{pq}r$ stands for the derivative of $r$ with respect to the edge-weight between nodes $p$ and $q$ (i.e., $A_{pq}=A_{qp}$). This closed-form gradient eliminates the need for numerical integration during the forward pass. Eq.~\eqref{nabla} can be computed in $\mathcal O(N^3)$ operations: we begin by pre-computing the parts independent of $p$ and $q$ indices as a matrix,
$
M_{ij}:=
\langle\boldsymbol{v}^i, \boldsymbol\omega\rangle
\langle\boldsymbol{v}^j, \boldsymbol\omega\rangle
/\left(\lambda_j^3 (1-\lambda_i/\lambda_j-\delta_{ij})\right)
$. 
Then we expand both parentheses of the form $(v^i_p-v^i_q)$ to obtain four terms of the form $\sum_{ij}v^i_p M_{ij}v^j_q$, which can be efficiently computed  as a product of three matrices.

\subsection*{Critical budget $b_c$ for global phase-locking}
Here, we present a simple argument to predict the critical budget $b_c$ given in Eq.~\eqref{b_c} corresponding to the onset of global phase-locking. Rewriting Eq.~\eqref{oscillator_eq} for the stationary state gives
\begin{equation}\label{stationary_eq_discrete}
\omega_i = \sum_{j=1}^{N} A_{ij} H(\theta_i, \theta_j).
\end{equation}
Next, we take the absolute value of both sides, sum over the index $i$, and apply the triangle inequality to obtain
\begin{equation}
\sum_i|\omega_{i}| \le \sum_{i,j} A_{ij} |H(\theta_i, \theta_j)|
\le \max(H) \sum_{i,j} A_{ij}.
\end{equation}
The last inequality also uses
$\max(H)\equiv \max_{\theta_i,\theta_j}|H(\theta_i,\theta_j)|$.
Using Eq.~\eqref{budget}, we rewrite this bound in terms of $b$ to obtain a lower bound on the budget giving rise to the globally phase-locked state
\begin{equation}\label{budget_inequality}
b \ge \frac{1}{\max(H)}\frac{1}{N}\sum_i|\omega_{i}|.
\end{equation}

Optimized networks are observed to saturate this bound (Fig.~\ref{fig:phase_diagrams}), which we accordingly identify as the critical budget $b_c$. At the conceptual level, the synchrony among oscillators is sustained by attractive interactions given by the product of the edge-weight $A_{ij}$ and the state-dependent function $H(\theta_i, \theta_j)$. Near the critical point, the coupling budget becomes highly constrained, compelling the optimization to identify a configuration that maximizes $H(\theta_i, \theta_j)$ across all connected node pairs.

Taking the thermodynamic limit $N\rightarrow\infty$, we replace the sum by integral $\tfrac{1}{N}\sum_i|\omega_{i}|\rightarrow\int{|\omega|g(\omega)d\omega}$. Additionally, our assumption of a mean-zero frequency distribution $g(\omega)$ implies   $\int_{-\infty}^0{(-\omega)g(\omega)d\omega} = \int_0^{\infty}{\omega g(\omega)d\omega}$. This results in a general prediction for $b_c$ given by
Eq.~\eqref{b_c}. The same result is derived in detail from a variational analysis for the Kuramoto model specifically Eq.~\eqref{eq:weak:bc}.

\subsection*{Derivation of the optimal pairing function $\nu(\omega)$}
Next, we derive the conditions~\eqref{bipartite} and \eqref{curve_eq} that determine the pairing function $\nu(\omega)$ for a supercritical budget, $b\ge b_c$, in the thermodynamic limit $N \rightarrow \infty$. Upon global phase-locking, each phase $\theta_i(t)$ asymptotically approaches a steady-state value, $\theta_i(t) \rightarrow \theta_i^*$. Prior to network construction, nodes are indistinguishable by any property other than their frequency. Thus, all node properties in the optimal configuration—including their positions in such a network, and final phases—can only depend on their frequencies and nothing else. This can be mathematically expressed as $A_{ij}\eqqcolon A(\omega_i,\omega_j)$ and $\theta_i^* \eqqcolon \theta^*(\omega_i) \eqqcolon \theta^*_{\omega_i}$.


With this notation, in the thermodynamic limit, we rewrite the stationary state equation~\eqref{stationary_eq_discrete} as
\begin{equation}\label{kuramoto_continuous}
\omega_i =N \int_{-\infty}^{\infty} g(\omega_j) A(\omega_i,\omega_j) H\big(\theta^*_{\omega_i},\theta^*_{\omega_j}\big)d\omega_j.
\end{equation}

We now incorporate into our analysis the key empirically observed property of   optimal networks: \emph{strong monophily}. To this end, we introduce a general ansatz for the adjacency matrix with such a property, ensuring that each node couples to only a narrow band of neighboring frequencies:
\begin{equation}\label{ansatz_A}
\begin{aligned}
A(\omega,\sigma) &= f(\omega,\sigma)\, e^{-\frac{1}{\epsilon}h(\omega,\sigma)}, \\
    h(\omega,\sigma) &\ge 0,\text{ with equality only when  }\sigma = \nu(\omega).
\end{aligned}
\end{equation}
Here, we have moved to a more convenient notation $(\omega_i,\omega_j)\rightarrow(\omega,\sigma)$. The variable $\epsilon\ll1$ controls how tightly the couplings $A(\omega,\sigma)$ concentrate around the curve $\sigma=\nu(\omega)$. Away from this curve, the positivity of $h$ leads to a strongly negative exponent, suppressing $A(\omega, \sigma)$ to negligible values. Functions $f(\omega,\sigma)$ and $h(\omega,\sigma)$ determine the variations of couplings along and transverse to the curve $\sigma=\nu(\omega)$, respectively Fig.~\ref{fig:adjacencies}(a). The undirected nature of the network implies symmetry $A(\omega,\sigma)=A(\sigma,\omega)$, which is ensured by demanding $f$ and $h$ to also be symmetric. Symmetry of $h$ further implies that $\nu$ must be a self-inverse, $\nu(\nu(\omega))=\omega$. 

Next, we utilize Laplace's method \cite{holmes2012introduction} to integrate Eq.~\eqref{kuramoto_continuous} with the ansatz~\eqref{ansatz_A}, yielding
\begin{equation}\label{omega}
    \omega = Ng(\nu) 
    H\big(\theta^*_\omega,\theta^*_\nu\big)
    A(\omega,\nu)\sqrt{\frac{2\pi\epsilon}{h_{\sigma\sigma}(\omega,\nu)}}.
\end{equation}
Here we have used the shorthand notations $\nu=\nu(\omega)$ and $h_{\sigma\sigma}(\omega,\nu)=\left.\frac{\partial^2h(\omega,\sigma)}{\partial \sigma^2}\right|_{\sigma=\nu(\omega)}$. Let us consider the same equation from the perspective of the target node with frequency $\nu(\omega)$. Since $\nu(\nu(\omega))=\omega$ we immediately get
\begin{equation}\label{nu}
    \nu = Ng(\omega) 
    H\big(\theta^*_\nu,\theta^*_\omega\big)
    A(\nu,\omega)\sqrt{\frac{2\pi\epsilon}{h_{\sigma\sigma}(\nu,\omega)}}.
\end{equation}
Using the antisymmetry of the coupling function $H(\theta,\phi)=-H(\phi,\theta)$, and symmetry of $A(\omega,\nu)=A(\nu,\omega)$, we can express the ratio of Eqs.~\eqref{omega} and \eqref{nu} as 
\begin{equation}\label{ratio}
    \frac{\omega}{\nu} = -\frac{g(\nu)}{g(\omega)} 
    \sqrt{\frac{h_{\sigma\sigma}(\nu,\omega)}{h_{\sigma\sigma}(\omega,\nu)}}.
\end{equation}

The final step involves simplifying the square root term. Note that  function $h$ obtains its minimum at $h(\omega,\nu(\omega))=0$. Hence both partial derivatives must vanish here: $h_\omega(\omega,\nu(\omega))=h_\sigma(\omega,\nu(\omega))=0$. This is true for all values of $\omega$, allowing us to differentiate further with respect to $\omega$: 
\begin{equation}
\begin{aligned}
    \frac{d}{d\omega}h_\omega(\omega,\nu(\omega)) 
    &= h_{\omega\omega}(\omega,\nu) + h_{\omega\sigma}(\omega,\nu)\nu'(\omega)=0, 
    \\
    \frac{d}{d\omega}h_\sigma(\omega,\nu(\omega)) 
    &= h_{\omega\sigma}(\omega,\nu) + h_{\sigma\sigma}(\omega,\nu)\nu'(\omega)=0.
\end{aligned}
\end{equation}
We solve this system of linear equations to obtain $ h_{\omega\omega}(\omega,\nu) = h_{\sigma\sigma}(\omega,\nu)\cdot(\nu'(\omega))^2$. Further, since $h(\omega,\sigma)=h(\sigma,\omega)$, we can change the differentiation variable and interchange the evaluation variable order yielding $h_{\omega\omega}(\omega,\nu)=h_{\sigma\sigma}(\nu,\omega)= h_{\sigma\sigma}(\omega,\nu)\cdot(\nu'(\omega))^2$.  Substitution into Eq.~\eqref{ratio} then yields
\begin{equation}\label{27}
    \frac{\omega}{\nu} = -\frac{g(\nu)}{g(\omega)} \left|\frac{d\nu}{d\omega}\right|.
\end{equation}
Note that the right-hand side is always negative, yielding a significant result: $\omega$ and $\nu(\omega)$ necessarily have opposite signs. In other words, the optimal network configuration is bipartite, with each node connecting to nodes with frequencies of opposite sign: $\omega \nu \le 0$. Allowing for both signs of $d\nu/d\omega$, the differential equation becomes $\nu g(\nu)\nu' \pm \omega g(\omega)=0$. Or putting it in a symmetric differential form, we find Eqs.~\eqref{bipartite} and \eqref{curve_eq}, the latter of which is reproduced here for convenience. 
 \begin{equation}\label{curve_eq_methods}
     \omega g(\omega) d\omega = \pm \nu g(\nu)d\nu.
 \end{equation}
 It should be mentioned that Eqs.~\eqref{27}--\eqref{curve_eq_methods} can be derived for alternative forms of the ansatz Eq.~\eqref{ansatz_A}, including the step-function and tent-shaped drop-offs of edge weights with increasing distance from the curve $\nu(\omega)$.

The boundary conditions for $\nu(\omega)$ in Eq.~\eqref{curve_eq_methods} can be found by first analyzing the sign   of the equation terms. Because $\omega$ and $\nu(\omega)$ have opposite signs and distribution $g(\omega)$ is non-negative, the differential terms must compensate for the sign discrepancy, leading to $d\nu_-/d\omega \ge 0$ and $d\nu_+/d\omega \le 0$. That is, $\nu_-$ is increasing and $\nu_+$ is decreasing. To achieve global phase-locking, any oscillator with frequency $\omega$ such that $|\omega| > 0$ must be connected to some other oscillator. For the decreasing branch, the largest positive frequency must   connect to the  largest negative frequency, $\nu_+(\omega_\text{max}) = \omega_\text{min}$, and smallest positive frequency connects to the smallest negative one, $\nu_+(0^+) = 0^-$. Similarly, for the increasing branch, we obtain $\nu_-(\omega_\text{max}) = 0^-$ and $\nu_-(0^-) = \omega_\text{max}$. Note that the bounds $\omega_\text{min}$ and $\omega_\text{max}$ can be either finite or extend to $\pm\infty$.

As a simple example, let us briefly discuss the uniformly distributed frequencies $g(\omega)=1/2$ for $-1\le\omega\le 1$. Equation \eqref{curve_eq} then reduces to $\omega^2\pm\nu^2=c$. The positive branch with the appropriate boundary condition  results in the trivial $\nu(\omega)_+=-\omega$. Such a network couples frequency pairs with equal magnitude and opposite signs. The more favorable negative branch has a more interesting structure $\nu_-=-\text{sign}(\omega)\sqrt{1-\omega^2}$.

\subsection*{Derivation of the optimal node strengths $s(\omega)$}

The strength $s_i=\sum_{j=1}^{N} A_{ij}$ of node $i$ is the summation of its edge weights, and it represents the share of the total budget $Nb$ allocated to $i$. In the thermodynamic limit, we set $s_i=s(\omega_i)$ using a strength function $s(\omega)$, allowing us to express the budget as
\begin{equation}\label{budget_strength}
b = \int 
    \underbrace{
        s(\omega)g(\omega)
    }_{\mathcal L_b(s, \omega)} 
    \, 
    d\omega.
\end{equation}
We denote the integrand by $\mathcal L_b(s,\omega)$ for convenience. Below we will show that given a fixed strength allocation $s(\omega)$, the optimal order parameter $r$ can be expressed as the following integral
\begin{equation} \label{r_pre_optimization}
    r=
    \int
    \underbrace{
        \frac{1}{2}
        \sqrt{1+\frac{\omega^2}{\nu^2}-2\frac{\omega}{\nu}\sqrt{1-\frac{\omega^2}{s(\omega)^2}}}
        \ g(\omega)
    }_{\mathcal L_r(s, \omega)} 
    \,
    d\omega.
\end{equation}
We denote the integrand by $\mathcal L_r(s,\omega)$. From this perspective, the optimization of synchrony naturally reduces to a constrained variational problem in which we seek to maximize $r$ for fixed $b$. That is, we look for the function $s(\omega)$ that maximizes Eq.~\eqref{r_pre_optimization} while satisfying Eq.~\eqref{budget_strength}. 

To proceed, we introduce a Lagrangian $\mathcal L=\mathcal L_r+\lambda\mathcal L_b$ with   Lagrange multiplier $\lambda$ and   derive the Euler–Lagrange equation \cite{calder2020calculus}
\begin{equation}
    \frac{\partial \mathcal L_r}{\partial s} + \lambda\frac{\partial \mathcal L_b}{\partial s} = 0.
\end{equation}
Note that we have omitted  terms of the form $\partial/\partial s'$,  since $\mathcal L_r$ and $\mathcal L_b$ do not depend on $s'(\omega)=ds(\omega)/d\omega$. Upon differentiation and simplification, we find
\begin{equation}
    \omega^3 g(\omega) = 4 \lambda \nu(\omega
    ) s^2 \sqrt{s^2-\omega^2}\mathcal L_r(s,\omega),
\end{equation}
which further simplifies to yield Eq.~\eqref{s_condition} (where  $c=4/\lambda^2$). (Both  $c$ and $\lambda$ enforce the budget constraint, and  so we also refer to $c$ as a Lagrange multiplier.)

The rest of this section is devoted to the derivation of Eq.~\eqref{r_pre_optimization}. We first obtain an expression for the order parameter $r$ that takes advantage of the pairing function $\nu(\omega)$. We begin by considering the form
$r=\tfrac{1}{N}\sum_j \cos\theta_j$, which uses the definition $re^{i\phi}=\tfrac{1}{N}\sum_j e^{i\theta_j}$ (equivalent to Eq.~\eqref{order_param}) and applies a change of variables $\theta_j\rightarrow\theta_j+\phi$ so that the angular mean is zero. In the thermodynamic limit, we can then write 
\begin{equation}\label{r_long1}
\begin{split}
r=&
\int_{\omega_\text{min}}^{\omega_\text{max}}\cos(\theta^*_\omega)g(\omega)d\omega
\\=&
\int_{\omega_\text{min}}^0 \frac{\cos(\theta^*_\omega)}{\omega}\omega g(\omega)d\omega
+
\int_0^{\omega_\text{max}} \frac{\cos(\theta^*_\omega)}{\omega}\omega g(\omega)d\omega.
\end{split}
\end{equation}
Next, we invoke Eq.~\eqref{curve_eq_methods} and change the variable of integration in the second integral, which introduces a $\pm$ sign in front of the second term. However, as seen in the last subsection, the choice of the pairing function branch (i.e., $\nu_-$ versus $\nu_+$) effectively interchanges the integration limits and cancels the $\pm$ sign to yield
\begin{equation}\label{r_long2}
\begin{split}
r=&
\int_{\omega_\text{min}}^0 \frac{\cos(\theta^*_\omega)}{\omega}\omega g(\omega)d\omega
\pm
\int_{\nu_\pm(0)}^{\nu_\pm(\omega_\text{max})} \frac{\cos(\theta^*_\omega)}{\omega}\nu g(\nu)d\nu
\\=&
\int_{\omega_\text{min}}^0 \frac{\cos(\theta^*_\omega)}{\omega}\omega g(\omega)d\omega
-
\int_{\omega_\text{min}}^0 \frac{\cos(\theta^*_\omega)}{\omega}\nu g(\nu)d\nu.
\end{split}
\end{equation}
Then, we exploit the self-inverse nature of the pairing function $\nu(\omega)$ to interchange the variable labels $\omega$ and $\nu$ in the second integral to obtain
\begin{equation}\label{r_long3}
\begin{split}
r=&
\int_{\omega_\text{min}}^0 \frac{\cos(\theta^*_\omega)}{\omega}\omega g(\omega)d\omega
-
\int_{\omega_\text{min}}^0 \frac{\cos(\theta^*_\nu)}{\nu}\omega g(\omega)d\omega
\\=&
\frac{1}{2}
\int_{\omega_\text{min}}^{\omega_\text{max}}
\underbrace{
    \left[
        \frac{\cos(\theta^*_\omega)}{\omega}
        -
        \frac{\cos(\theta^*_\nu)}{\nu}
    \right] 
}_{\chi}
\omega g(\omega)d\omega.
\end{split}
\end{equation}
Here, the second line combines the two integral terms and extends the domain to positive frequencies, introducing a factor of $1/2$ to compensate for the doubling. The validity of this last step can be readily confirmed by applying the same sequence of transformations to the first term on line two of Eq.~\eqref{r_long1} instead of the second. The bracketed term $\chi$ in Eq.~\eqref{r_long3} accounts for contributions from connected node pairs $\omega$ and $\nu(\omega)$, and the integration measure $\omega g(\omega)d\omega$ is invariant, up to a sign, under the variable exchange $\omega\leftrightarrows\nu$. 

Next, we aim to express $r$ as an explicit function of the strength function $s(\omega)$ rather than stationary phases $\theta_\omega^*$. To this end, we express $s(\omega)$ in the thermodynamic limit as $s(\omega)=N\int g(\sigma)A(\omega,\sigma)d\sigma$. Assuming the strong-monophily (e.g., as indicated by the ansatz given by Eq.~\eqref{ansatz_A}), we approximate the integral to leading order using Laplace's method and obtain
$s(\omega)=Ng(\nu)A(\omega,\nu)\sqrt{{2\pi\epsilon}/{h_{\sigma\sigma}(\omega,\nu)}}$.
Combining this result with Eq.~\eqref{omega} yields $\omega = s(\omega)\sin{(\theta^*_\omega-\theta^*_\nu)}$, or equivalently
\begin{equation}\label{phase_gap}
    \theta^*_\omega-\theta^*_\nu = \arcsin\frac{\omega}{s(\omega)}\eqqcolon C_\omega.
\end{equation}
That is, in an optimal network the phase gap between two oscillators is precisely determined by an oscillator's frequency $\omega$ and node strength $s(\omega)$. By  symmetry, one also has for any neighboring node that $C_\nu=-C_\omega$. 

Next, we use Eq.~\eqref{phase_gap} to further study the bracketed term in Eq.~\eqref{r_long3}. We find 
\begin{equation}\label{bracketed_term1}
\begin{split}
    \chi &= \frac{\cos(\theta^*_\omega)}{\omega}
    -
    \frac{\cos(\theta^*_\nu)}{\nu}
    \\&=
    \frac{\cos(\theta^*_\omega)}{\omega}
    -
    \frac{\cos(\theta^*_\omega-C_\omega)}{\nu}
    \\&=
    \frac{\sin C_\omega}{\nu}\sin(\theta^*_\omega) 
    +
    \left(\frac{1}{\omega}-\frac{\cos C_\omega}{\nu} \right)\cos(\theta^*_\omega)
    \\&\eqqcolon
    \ p\sin(\theta^*_\omega) 
    +
    q \cos(\theta^*_\omega),
\end{split}
\end{equation}
where we have defined the coefficients of $\sin(\theta^*_\omega)$ and $\cos(\theta^*_\omega)$ as $p$ and $q$, respectively. Next, we factor out the common amplitude $\sqrt{p^2 + q^2}$ and express the remaining coefficients in terms of $\phi_\omega \coloneqq \arcsin\left(p/\sqrt{p^2+q^2}\right)$:
\begin{equation}\label{bracketed_term2}
\begin{split}
    \chi
    &=
    \sqrt{p^2+q^2}\left(
    \sin\phi_\omega\sin(\theta^*_\omega) 
    +
    \cos\phi_\omega \cos(\theta^*_\omega)
    \right)
    \\&=
    \sqrt{p^2+q^2}
    \cos(\theta^*_\omega-\phi_\omega)
    \\&=
    \sqrt{\frac{1}{\nu^2}+\frac{1}{\omega^2}-\frac{2}{\omega\nu}\sqrt{1-\frac{\omega^2}{s(\omega)^2}}}
    \cos(\theta^*_\omega-\phi_\omega).
\end{split}
\end{equation}
In the last line, we substituted the definitions of $p$, $q$, and $C_\omega$ back into the expression and simplified. 

Each pair of connected nodes with frequencies  $\omega$ and $\nu$ introduces two ``phase'' degrees of freedom:  $\theta_\omega^*$ and $\theta_\nu^*$. One of these phases was fixed by the phase-gap equation (i.e., Eq.~\eqref{phase_gap}), which we applied in Eq.~\eqref{bracketed_term2}. The other, $\theta_\omega^*$, remains as an explicit variable. The system’s optimality requires  each degree of freedom to be adjusted so as to maximize $r$, and for Eq.~\eqref{bracketed_term2}, this is achieved when $\cos(\theta_\omega^*-\phi_\omega)=1$, or $\theta_\omega^* = \phi_\omega$. (In the next section, we use this relation to explicitly solve for the stationary phases.)
We then substitute this relation into Eq.~\eqref{bracketed_term2}, and it follows that
Eq.~\eqref{r_long3} 
recovers 
Eq.\eqref{r_pre_optimization}.

\subsection*{Asymptotic derivations}
Considering Eq.~\eqref{s_condition}, observe that due to the term under the square root,   a real-valued solution is admitted only when the node strength satisfies $s(\omega)>|\omega|$ (which is a necessary condition for the system to be in the supercritical regime with $b\ge b_c$). Here, we derive asymptotic expressions for when $s(\omega)>|\omega|$ but $s(\omega)\approx|\omega|$.
%
%
%
To this end, we adopt the notation $1-\omega^2/s^2 \eqqcolon \epsilon(\omega) \ll 1$. Leading order expansion of Eq.~\eqref{s_condition} yields the optimal distribution of node strengths $s(\omega)\approx|\omega|\left(1+(\omega^2+\nu^2)^{-1}/c\right)$. We compute $c$ by enforcing the budget constraint given by Eq.~\eqref{budget_strength}, which is also used to reveal the critical budget $b_c$. Finally, we insert this expression for $s(\omega)$ into Eq.~\eqref{r_pre_optimization} and similarly Taylor expand to obtain the asymptotic expansion given in Eqs.~\eqref{eq:weak:r}-\eqref{eq:weak:bc}.

In the strong coupling regime, we assume $b$ is sufficiently large such that each node receives a strength allocation $s(\omega) \gg |\omega|$. Under this assumption, Eq.~\eqref{s_condition} simplifies considerably in the leading-order approximation, yielding an explicit expression for the optimal strength as given in Eq.~\eqref{eq:strongly_coupled:s}.
This analytic form allows direct evaluation of $r$. Substituting Eq.~\eqref{eq:strongly_coupled:s} into Eq.~\eqref{r_pre_optimization}, and expanding the integrand consistently within the same asymptotic limit, we obtain Eq.~\eqref{eq:strongly_coupled:r}.

Finally, we derive the explicit form of the stationary phases $\theta_\omega^*$. Observing that the cosine term in Eq.~\eqref{bracketed_term2} is maximized when $\theta^*(\omega) = \phi_\omega$, where $\sin \phi_\omega = p / \sqrt{p^2 + q^2}$, we apply the same asymptotic methods as above and obtain Eq.~\eqref{eq:strongly_coupled:theta}.

\subsection*{Proof for the absence of a synchronization threshold}
\label{sec:proof}
Here, we prove that optimized oscillator networks lack a threshold for synchronization (i.e., lack an incoherent phase) in the thermodynamic, large-$N$ limit. That is, a nonzero order parameter  occurs  for any nonzero budget: $b>0 \implies r>0$. Our approach relies on considering a specific allocation of the small budget in which the full budget $Nb$ (recall $b$ is a per-node budget) is concentrated onto a small subset of oscillators, and the set of $\tilde N\ll N$ oscillators is chosen to be sufficiently small so that oscillators within the group phase-lock (whereas oscillators outside the group remain drifting). Focusing on $r\approx0$, we study the $N\to\infty$ behavior of $r$ for this particular configuration, which lower bounds $r$ for any optimal configuration.

Let us first assume that $0<b\ll1$. We consider synchronization for a small set of oscillators with mean frequency $\omega$, which we set to be zero without loss of generality.
For notational simplicity, we further assume that the distribution of frequencies satisfies $g(0)>0$ and  $g'(0)=0$. To allocate the edge-weight budget $b$, we will consider a thin, width-$\Delta\omega\ll1$  slice of the frequency band, $\omega\in\Omega\equiv [-\Delta\omega/2,\Delta\omega/2]$, and spend the full budget $Nb$ on this small group of oscillators to ensure that they phase-lock. To   leading order, the number of oscillators in that group is $\tilde N=Ng(0)\Delta\omega$. If we concentrate the full budget allowance $Nb$ on these $\tilde N$ nodes, then the effective per-node budget is given by $\tilde b=b/\left(g(0)\Delta\omega\right)$. The effective frequency distribution within this target group to the leading order is $\tilde g(\omega)=1/\Delta\omega$ for $\omega\in \Omega$, and Eq.~\eqref{b_c} allows us to compute the effective critical budget for that group to be $\tilde b_c=\Delta\omega/4$. Therefore, for any per-node budget $b$, the subgroup of $\tilde{N}\ll N$ oscillators is ensured to phase-lock when $\Delta\omega \le\sqrt{4b/g(0)}$.  


To proceed, we set $\Delta\omega$ to this bound and calculate the associated order parameter across this subgroup using Eq.~\eqref{eq:weak:rc}. 
Following the same derivation as before for the uniform frequency distributions, we get $\nu_-^2=\Delta \omega^2/4-\omega^2$, and integrating Eq.~\eqref{eq:weak:rc} gives us $\tilde r_c=\pi/4$.

Finally, because all oscillators outside the target subgroup have remained uncoupled and are drifting,
they do not contribute to the order parameter in the thermodynamic limit.
It follows that the global order parameter $r$ is given by 
\begin{equation}
    \label{r_bound}
    r=\frac{\tilde N}{N} \tilde r_c = g(0) \Delta \omega \frac{\pi}{4} =\frac{\pi}{2} \sqrt{b g(0)}
\end{equation}
This shows that even for an extremely small positive budget $b$, an optimal network's order parameter is bounded below by Eq.~\eqref{r_bound}, and therefore it is greater than zero even in the thermodynamic limit (i.e., $b>0\implies r>0$). Note that the assumptions on $g(\omega)$ were made purely for convenience and can be relaxed by choosing a frequency band around an arbitrary point $\omega_0$ with $g(\omega_0)>0$. And even though $g'(\omega_0)$ may be nonzero, we can still sample oscillators with uniform frequency distribution $\tilde g(\omega)$ in the selected frequency band to obtain the same result.

\subsection*{Experiment description for other synchronizing systems}

To assess how broadly the structural findings in our paper arise across different settings, we numerically optimized networks for several oscillatory dynamical systems, as described in the Results. Here we provide details of the dynamical processes and optimization targets. The dynamics of the Sakaguchi–Kuramoto model \cite{sakaguchi1986soluble} are given by
$$
\dot\theta_i = \omega_i + \sum_j A_{ij}\sin(\theta_j - \theta_i + \phi).
$$
This matches the Kuramoto model with the only distinction being the phase lag $\phi$. The optimization target is still the Kuramoto order parameter in Eq.~\eqref{order_param}. We use $\phi=0.1$, and note that for much larger phase lag, the straightforward optimization ceases to converge to the expected network structure.

The power grid swing equations \cite{filatrella2008analysis} are given by
$$
\ddot\theta_i + \dot\theta_i = \omega_i + \sum_j A_{ij}\sin(\theta_j - \theta_i),
$$
which incorporates inertia via a second-order derivative. The synchrony is again measured by the Kuramoto order parameter Eq.~\eqref{order_param}. Note that equilibrium for this equation is identical to that for the Kuramoto model, and the synchrony alignment function can be used to accelerate numerical optimization.

The Stuart--Landau model \cite{matthews1990phase} is the normal form of coupled phase--amplitude oscillators near the Hopf bifurcation 
$$
\dot z_j = (1 + i\omega_j - (1+i)|z_j|^2)z_j + \sum_k A_{jk}(z_k - z_j).
$$
Here $z_j$ is the complex state of node $j$, admitting the standard phase--amplitude decomposition $z_j(t) = \rho_j(t) e^{i \theta_j(t)}$. The node phases $\theta_j(t)$ are then used with Eq.~\eqref{order_param} to compute the synchronization order parameter. 

For the three systems above, we have sampled the frequencies $\{\omega_i\}$ uniformly from the range $(-0.5,0.5)$. The initial conditions are chosen as $\theta_i(0)=0$ for each model, and in the case of the Stuart–Landau model, we also set $\rho_i(0)=1$.

Finally, the coupled R\"ossler oscillator dynamics \cite{leyva2012explosive,skardal2017optimal} are given by the coupled chaotic equations
\begin{equation}
\begin{aligned}
&\dot{x}_i = -\omega_i \left[ \gamma \left( x_i - \sum_{j=1}^{N} A_{ij}(x_j - x_i) \right) + \beta y_i + \lambda z_i \right],\\
&\dot{y}_i = -\omega_i \left( -x_i + \nu y_i \right),\\
&\dot{z}_i = -\omega_i \left[ -g(x_i) + z_i \right],\\
&g(x) =
    \begin{cases}
    0, & \text{if } x \le 3, \\
    \mu (x - 3), & \text{if } x > 3.
    \end{cases}
\end{aligned}
\end{equation}
The non-linearity in the equation for $\dot z_i$ has been modified from the original R\"ossler equations in order to simplify the  electronic experiments \cite{leyva2012explosive}.
An appropriate scalar measure for oscillator phases for this model is the rotation angle around the $z$ axis (i.e., the azimuthal angle). With these phases, we compute and optimize the order parameter Eq.~\eqref{order_param}. The frequencies are uniformly sampled from the range $(9.8,10.2)$. Other parameters are $\gamma=0.05$, $\beta=0.5$, $\lambda=1$, $\mu=15$, and $\nu=-0.08$. The optimization experiments for these synchronizing systems rely on the GradNet framework developed in \cite{mikaberidze2026gradnet}.

\edit{
\subsection*{Power grid optimization with length-proportional costs}

We use PanTaGruEl, an open-access model of the continental European synchronous transmission grid that combines realistic geography, line parameters, and generation dispatch at continental scale. Following standard phase-oscillator representations of power-grid dynamics \cite{filatrella2008analysis,pagnier2019inertia}, we construct a reduced oscillator network from the first supplied operating snapshot of the PanTaGruEl dataset \cite{pagnier_2022_5816149}. We contract buses with identical geographic coordinates, summing their net active-power injections and the susceptance magnitudes of resulting parallel branches and removing resulting self-loops. We then center the aggregated injections and divide both the injections and coupling weights by the injection standard deviation, yielding unit-variance oscillator frequencies $\omega_i$. We do not apply an additional operating-voltage-magnitude factor to the supplied PanTaGruEl susceptances. The resulting network contains $3,258$ nodes and $4,364$ edges.


We generalize the coupling budget to include edge-dependent costs that grow linearly with edge length. Each edge is assigned a cost

$$
c_{\sigma}(i,j) = (1-\sigma) + \sigma d_{ij}, \qquad 0\leq\sigma\leq1,
$$
where $d_{ij}=D_{ij}/\bar D$, with $D_{ij}$ the great-circle distance between the endpoints and $\bar D$ the mean great-circle distance over all retained node pairs. Thus, $\sigma=0$ assigns equal cost per unit coupling to every edge, whereas $\sigma=1$ gives length-proportional costs. 
For each value of $\sigma$, we define the empirical cost budget
$$
b_\sigma=\frac{1}{N}\sum_{ij}c_\sigma(i,j)A^{\rm real}_{ij},
$$
and hold this budget fixed during optimization,
$\tfrac{1}{N}\sum_{ij}c_\sigma(i,j)A_{ij}=b_\sigma$.

To ask how the empirical grid’s existing coupling would be reallocated if synchrony were the sole objective, we keep the observed topology fixed and optimize only the weights of existing edges. This restriction also keeps the optimization computationally tractable: allowing arbitrary new connections would increase the number of candidate edges from $4,364$ to more than $5\times10^6$. 

For the trajectory shown in Fig.~\ref{fig:pantagruel}(b), the initial edge weights are generated by drawing independent exponential random variables and normalizing them to obtain random shares of the empirical cost budget.

To accelerate optimization at this system size, we use the strong-synchrony SAF approximation and its analytical edge-weight gradient described above in Eqs.~\eqref{saf}-\eqref{nabla}, rather than differentiating through the full Kuramoto dynamics. 
The reported $r$ values are terminal order parameters from direct Kuramoto simulations initialized with zero phases, with relaxation assessed after $50$ time units and simulations terminated upon satisfying the relaxation criterion or after $1,000$ time units.
}

\section*{Acknowledgements}
This work is supported by the U.S. National Science Foundation under Grant No. DMS-2401276. Any opinions, findings, and conclusions or recommendations expressed in this material are those of the authors and do not necessarily reflect the views of the NSF. The authors also acknowledge funding from the University of Wyoming including the AI Matching Funds program, an AI seed grant, and the Center for Wildlife, Technology and Computing (WyldTech).

\bibliographystyle{unsrtnat}
\bibliography{references}

@article{kuramoto1975international,
  title={International symposium on mathematical problems in theoretical physics},
  author={Kuramoto, Yoshiki},
  journal={Lecture Notes in Physics},
  volume={30},
  pages={420},
  year={1975}
}

@book{pikovsky2001universal,
  title={Synchronization, A Universal Concept in Nonlinear Sciences},
  author={Pikovsky, Arkady and Rosenblum, Michael and Kurths, J{\"u}rgen},
  year={2001},
  publisher={Cambridge University Press}
}

@article{strogatz2000kuramoto,
  title={From {K}uramoto to {C}rawford: exploring the onset of synchronization in populations of coupled oscillators},
  author={Strogatz, Steven H},
  journal={Physica D: Nonlinear Phenomena},
  volume={143},
  number={1-4},
  pages={1--20},
  year={2000},
  publisher={Elsevier}
}

@article{chamlagai2022grass,
  title={Grass-roots optimization of coupled oscillator networks},
  author={Chamlagai, Pranick R and Taylor, Dane and Skardal, Per Sebastian},
  journal={Physical Review E},
  volume={106},
  number={3},
  pages={034202},
  year={2022},
  publisher={APS}
}

@article{arenas2008synchronization,
  title={Synchronization in complex networks},
  author={Arenas, Alex and D{\'\i}az-Guilera, Albert and Kurths, J\"urgen and Moreno, Yamir and Zhou, Changsong},
  journal={Physics Reports},
  volume={469},
  number={3},
  pages={93--153},
  year={2008},
  publisher={Elsevier}
}

@article{rodrigues2016kuramoto,
  title={The {K}uramoto model in complex networks},
  author={Rodrigues, Francisco A and Peron, Thomas K DM and Ji, Peng and Kurths, J{\"u}rgen},
  journal={Physics Reports},
  volume={610},
  pages={1--98},
  year={2016},
  publisher={Elsevier}
}

@article{restrepo2005onset,
  title={Onset of synchronization in large networks of coupled oscillators},
  author={Restrepo, Juan G and Ott, Edward and Hunt, Brian R},
  journal={Physical Review E—Statistical, Nonlinear, and Soft Matter Physics},
  volume={71},
  number={3},
  pages={036151},
  year={2005},
  publisher={APS}
}

@article{pecora1998master,
  title={Master stability functions for synchronized coupled systems},
  author={Pecora, Louis M and Carroll, Thomas L},
  journal={Physical Review Letters},
  volume={80},
  number={10},
  pages={2109},
  year={1998},
  publisher={APS}
}

@article{barahona2002synchronization,
  title={Synchronization in small-world systems},
  author={Barahona, Mauricio and Pecora, Louis M},
  journal={Physical Review Letters},
  volume={89},
  number={5},
  pages={054101},
  year={2002},
  publisher={APS}
}

@article{nishikawa2006synchronization,
  title={Synchronization is optimal in nondiagonalizable networks},
  author={Nishikawa, Takashi and Motter, Adilson E},
  journal={Physical Review E},
  volume={73},
  number={6},
  pages={065106},
  year={2006},
  publisher={APS}
}

@article{donetti2005entangled,
  title={Entangled networks, synchronization, and optimal network topology},
  author={Donetti, Luca and Hurtado, Pablo I and Mu\~noz, Miguel A},
  journal={Physical Review Letters},
  volume={95},
  number={18},
  pages={188701},
  year={2005},
  publisher={APS}
}

@article{sorrentino2008adaptive,
  title={Adaptive synchronization of dynamics on evolving complex networks},
  author={Sorrentino, Francesco and Ott, Edward},
  journal={Physical Review Letters},
  volume={100},
  number={11},
  pages={114101},
  year={2008},
  publisher={APS}
}

@article{nishikawa2010network,
  title={Network synchronization landscape reveals compensatory structures, quantization, and the positive effect of negative interactions},
  author={Nishikawa, Takashi and Motter, Adilson E},
  journal={Proceedings of the National Academy of Sciences},
  volume={107},
  number={23},
  pages={10342--10347},
  year={2010},
  publisher={National Academy of Sciences}
}

@article{skardal2014optimal,
  title={Optimal synchronization of complex networks},
  author={Skardal, Per Sebastian and Taylor, Dane and Sun, Jie},
  journal={Physical Review Letters},
  volume={113},
  number={14},
  pages={144101},
  year={2014},
  publisher={APS}
}

@article{taylor2016synchronization,
  title={Synchronization of heterogeneous oscillators under network modifications: Perturbation and optimization of the synchrony alignment function},
  author={Taylor, Dane and Skardal, Per Sebastian and Sun, Jie},
  journal={SIAM Journal on Applied Mathematics},
  volume={76},
  number={5},
  pages={1984--2008},
  year={2016},
  publisher={SIAM}
}

@article{skardal2016optimal,
  title={Optimal synchronization of directed complex networks},
  author={Skardal, Per Sebastian and Taylor, Dane and Sun, Jie},
  journal={Chaos: An Interdisciplinary Journal of Nonlinear Science},
  volume={26},
  number={9},
  year={2016},
  publisher={AIP Publishing}
}

@article{skardal2019synchronization,
  title={Synchronization of network-coupled oscillators with uncertain dynamics},
  author={Skardal, Per Sebastian and Taylor, Dane and Sun, Jie},
  journal={SIAM Journal on Applied Mathematics},
  volume={79},
  number={6},
  pages={2409--2433},
  year={2019},
  publisher={SIAM}
}

@article{chen2018neuralode,
  title={Neural Ordinary Differential Equations},
  author={Chen, Ricky T. Q. and Rubanova, Yulia and Bettencourt, Jesse and Duvenaud, David},
  journal={Advances in Neural Information Processing Systems},
  year={2018}
}

@article{paszke2017automatic,
  title={Automatic differentiation in {PyTorch}},
  author={Paszke, Adam and Gross, Sam and Chintala, Soumith and Chanan, Gregory and Yang, Edward and DeVito, Zachary and Lin, Zeming and Desmaison, Alban and Antiga, Luca and Lerer, Adam},
  year={2017}
}

@misc{torchdiffeq,
	author={Chen, Ricky T. Q.},
	title={torchdiffeq},
	year={2018},
	url={https://github.com/rtqichen/torchdiffeq},
}

@article{ricci2021kuranet,
  title={KuraNet: systems of coupled oscillators that learn to synchronize},
  author={Ricci, Matthew and Jung, Minju and Zhang, Yuwei and Chalvidal, Mathieu and Soni, Aneri and Serre, Thomas},
  journal={arXiv preprint arXiv:2105.02838},
  year={2021}
}

@article{buzna2009synchronization,
  title={Synchronization in symmetric bipolar population networks},
  author={Buzna, Lubos and Lozano, Sergi and D{\'\i}az-Guilera, Albert},
  journal={Physical Review E—Statistical, Nonlinear, and Soft Matter Physics},
  volume={80},
  number={6},
  pages={066120},
  year={2009},
  publisher={APS}
}

@article{taylor2015topological,
  title={Topological data analysis of contagion maps for examining spreading processes on networks},
  author={Taylor, Dane and Klimm, Florian and Harrington, Heather A and Kram{\'a}r, Miroslav and Mischaikow, Konstantin and Porter, Mason A and Mucha, Peter J},
  journal={Nature Communications},
  volume={6},
  number={1},
  pages={7723},
  year={2015},
  publisher={Nature Publishing Group UK London}
}

@article{nishikawa2003heterogeneity,
  title={Heterogeneity in oscillator networks: Are smaller worlds easier to synchronize?},
  author={Nishikawa, Takashi and Motter, Adilson E and Lai, Ying-Cheng and Hoppensteadt, Frank C},
  journal={Physical Review Letters},
  volume={91},
  number={1},
  pages={014101},
  year={2003},
  publisher={APS}
}

@article{zhao2006relations,
  title={Relations between average distance, heterogeneity and network synchronizability},
  author={Zhao, Ming and Zhou, Tao and Wang, Bing-Hong and Yan, Gang and Yang, Hui-Jie and Bai, Wen-Jie},
  journal={Physica A: Statistical Mechanics and its Applications},
  volume={371},
  number={2},
  pages={773--780},
  year={2006},
  publisher={Elsevier}
}

@article{brede2008synchrony,
  title={Synchrony-optimized networks of non-identical Kuramoto oscillators},
  author={Brede, Markus},
  journal={Physics Letters A},
  volume={372},
  number={15},
  pages={2618--2622},
  year={2008},
  publisher={Elsevier}
}

@article{grabow2010small,
  title={Do small worlds synchronize fastest?},
  author={Grabow, Carsten and Hill, Steven M and Grosskinsky, Stefan and Timme, Marc},
  journal={Europhysics Letters},
  volume={90},
  number={4},
  pages={48002},
  year={2010},
  publisher={IOP Publishing}
}

@article{hong2002synchronization,
  title={Synchronization on small-world networks},
  author={Hong, Hyunsuk and Choi, Moo-Young and Kim, Beom Jun},
  journal={Physical Review E},
  volume={65},
  number={2},
  pages={026139},
  year={2002},
  publisher={APS}
}

@article{altenburger2018monophily,
  title={Monophily in social networks introduces similarity among friends-of-friends},
  author={Altenburger, Kristen M and Ugander, Johan},
  journal={Nature Human Behaviour},
  volume={2},
  number={4},
  pages={284--290},
  year={2018},
  publisher={Nature Publishing Group UK London}
}

@article{evtushenko2021paradox,
  title={The paradox of second-order homophily in networks},
  author={Evtushenko, Anna and Kleinberg, Jon},
  journal={Scientific Reports},
  volume={11},
  number={1},
  pages={13360},
  year={2021},
  publisher={Nature Publishing Group UK London}
}

@book{holmes2012introduction,
  title={Introduction to perturbation methods},
  author={Holmes, Mark H},
  volume={20},
  year={2012},
  publisher={Springer Science \& Business Media}
}

@article{kelly2011topology,
  title={On the topology of synchrony optimized networks of a {K}uramoto-model with non-identical oscillators},
  author={Kelly, David and Gottwald, Georg A},
  journal={Chaos: An Interdisciplinary Journal of Nonlinear Science},
  volume={21},
  number={2},
  year={2011},
  publisher={AIP Publishing}
}

@article{pinto2015optimal,
  title={Optimal synchronization of Kuramoto oscillators: A dimensional reduction approach},
  author={Pinto, Rafael S and Saa, Alberto},
  journal={Physical Review E},
  volume={92},
  number={6},
  pages={062801},
  year={2015},
  publisher={APS}
}

@article{brede2018competitive,
  title={Competitive influence maximization and enhancement of synchronization in populations of non-identical Kuramoto oscillators},
  author={Brede, Markus and Stella, Massimo and Kalloniatis, Alexander C},
  journal={Scientific Reports},
  volume={8},
  number={1},
  pages={702},
  year={2018},
  publisher={Nature Publishing Group UK London}
}

@article{papadopoulos2017development,
  title={Development of structural correlations and synchronization from adaptive rewiring in networks of Kuramoto oscillators},
  author={Papadopoulos, Lia and Kim, Jason Z and Kurths, J{\"u}rgen and Bassett, Danielle S},
  journal={Chaos: An Interdisciplinary Journal of Nonlinear Science},
  volume={27},
  number={7},
  year={2017},
  publisher={AIP Publishing}
}

@article{schafer2022understanding,
  title={Understanding Braess’ paradox in power grids},
  author={Sch{\"a}fer, Benjamin and Pesch, Thiemo and Manik, Debsankha and Gollenstede, Julian and Lin, Guosong and Beck, Hans-Peter and Witthaut, Dirk and Timme, Marc},
  journal={Nature Communications},
  volume={13},
  number={1},
  pages={5396},
  year={2022},
  publisher={Nature Publishing Group UK London}
}

@article{motter2013spontaneous,
  title={Spontaneous synchrony in power-grid networks},
  author={Motter, Adilson E and Myers, Seth A and Anghel, Marian and Nishikawa, Takashi},
  journal={Nature Physics},
  volume={9},
  number={3},
  pages={191--197},
  year={2013},
  publisher={Nature Publishing Group UK London}
}

@article{witthaut2012braess,
  title={Braess's paradox in oscillator networks, desynchronization and power outage},
  author={Witthaut, Dirk and Timme, Marc},
  journal={New journal of physics},
  volume={14},
  number={8},
  pages={083036},
  year={2012},
  publisher={IOP Publishing}
}

@article{wang2016enhancing,
  title={Enhancing synchronization stability in a multi-area power grid},
  author={Wang, Bing and Suzuki, Hideyuki and Aihara, Kazuyuki},
  journal={Scientific Reports},
  volume={6},
  number={1},
  pages={26596},
  year={2016},
  publisher={Nature Publishing Group UK London}
}

@article{romera2018vowel,
  title={Vowel recognition with four coupled spin-torque nano-oscillators},
  author={Romera, Miguel and Talatchian, Philippe and Tsunegi, Sumito and Abreu Araujo, Flavio and Cros, Vincent and Bortolotti, Paolo and Trastoy, Juan and Yakushiji, Kay and Fukushima, Akio and Kubota, Hitoshi and others},
  journal={Nature},
  volume={563},
  number={7730},
  pages={230--234},
  year={2018},
  publisher={Nature Publishing Group UK London}
}

@article{romera2022binding,
  title={Binding events through the mutual synchronization of spintronic nano-neurons},
  author={Romera, Miguel and Talatchian, Philippe and Tsunegi, Sumito and Yakushiji, Kay and Fukushima, Akio and Kubota, Hitoshi and Yuasa, Shinji and Cros, Vincent and Bortolotti, Paolo and Ernoult, Maxence and others},
  journal={Nature Communications},
  volume={13},
  number={1},
  pages={883},
  year={2022},
  publisher={Nature Publishing Group UK London}
}

@article{torrejon2017neuromorphic,
  title={Neuromorphic computing with nanoscale spintronic oscillators},
  author={Torrejon, Jacob and Riou, Mathieu and Araujo, Flavio Abreu and Tsunegi, Sumito and Khalsa, Guru and Querlioz, Damien and Bortolotti, Paolo and Cros, Vincent and Yakushiji, Kay and Fukushima, Akio and others},
  journal={Nature},
  volume={547},
  number={7664},
  pages={428--431},
  year={2017},
  publisher={Nature Publishing Group UK London}
}

@article{antonio2012frequency,
  title={Frequency stabilization in nonlinear micromechanical oscillators},
  author={Antonio, Dario and Zanette, Dami{\'a}n H and L{\'o}pez, Daniel},
  journal={Nature Communications},
  volume={3},
  number={1},
  pages={806},
  year={2012},
  publisher={Nature Publishing Group UK London}
}

@article{olfati2007consensus,
  title={Consensus and cooperation in networked multi-agent systems},
  author={Olfati-Saber, Reza and Fax, J Alex and Murray, Richard M},
  journal={Proceedings of the IEEE},
  volume={95},
  number={1},
  pages={215--233},
  year={2007},
  publisher={IEEE}
}

@article{kia2019tutorial,
  title={Tutorial on dynamic average consensus: The problem, its applications, and the algorithms},
  author={Kia, Solmaz S and Van Scoy, Bryan and Cort\'es, Jorge and Freeman, Randy A and Lynch, Kevin M and Mart\'{\i}nez, Sonia},
  journal={IEEE Control Systems Magazine},
  volume={39},
  number={3},
  pages={40--72},
  year={2019},
  publisher={IEEE}
}

@article{kamthe2025external,
  title={External Bias and Opinion Clustering in Cooperative Networks},
  author={Kamthe, Akshay Nagesh and Thota, Vishnudatta and Shrinate, Aashi and Tripathy, Twinkle},
  journal={Automatica},
  volume={175},
  pages={112224},
  year={2025},
  publisher={Elsevier}
}

@software{mikaberidze2025networkoptimization,
  author       = {Guram Mikaberidze},
  title        = {Network Optimization for Synchrony},
  year         = {2025},
  version      = {commit at 2025-04-17},
  url          = {https://gitlab.com/mikaberidze/network-optimization-for-synchrony},
  note         = {GitLab repository}
}

@article{daido1990intrinsic,
  title={Intrinsic fluctuations and a phase transition in a class of large populations of interacting oscillators},
  author={Daido, Hiroaki},
  journal={Journal of Statistical Physics},
  volume={60},
  number={5},
  pages={753--800},
  year={1990},
  publisher={Springer}
}

@article{rohden2014impact,
  title={Impact of network topology on synchrony of oscillatory power grids},
  author={Rohden, Martin and Sorge, Andreas and Witthaut, Dirk and Timme, Marc},
  journal={Chaos: An Interdisciplinary Journal of Nonlinear Science},
  volume={24},
  number={1},
  year={2014},
  publisher={AIP Publishing}
}

@article{skardal2011cluster,
  title={Cluster synchrony in systems of coupled phase oscillators with higher-order coupling},
  author={Skardal, Per Sebastian and Ott, Edward and Restrepo, Juan G},
  journal={Physical Review E—Statistical, Nonlinear, and Soft Matter Physics},
  volume={84},
  number={3},
  pages={036208},
  year={2011},
  publisher={APS}
}

@article{komarov2013multiplicity,
  title={Multiplicity of singular synchronous states in the Kuramoto model of coupled oscillators},
  author={Komarov, Maxim and Pikovsky, Arkady},
  journal={Physical Review Letters},
  volume={111},
  number={20},
  pages={204101},
  year={2013},
  publisher={APS}
}

@article{timme2006speed,
  title={Speed of synchronization in complex networks of neural oscillators: analytic results based on random matrix theory},
  author={Timme, Marc and Geisel, Theo and Wolf, Fred},
  journal={Chaos: An Interdisciplinary Journal of Nonlinear Science},
  volume={16},
  number={1},
  year={2006},
  publisher={AIP Publishing}
}

@article{witthaut2022collective,
  title={Collective nonlinear dynamics and self-organization in decentralized power grids},
  author={Witthaut, Dirk and Hellmann, Frank and Kurths, J{\"u}rgen and Kettemann, Stefan and Meyer-Ortmanns, Hildegard and Timme, Marc},
  journal={Reviews of Modern Physics},
  volume={94},
  number={1},
  pages={015005},
  year={2022},
  publisher={APS}
}

@article{menck2013basin,
  title={How basin stability complements the linear-stability paradigm},
  author={Menck, Peter J and Heitzig, Jobst and Marwan, Norbert and Kurths, J{\"u}rgen},
  journal={Nature Physics},
  volume={9},
  number={2},
  pages={89--92},
  year={2013},
  publisher={Nature Publishing Group UK London}
}

@article{wiley2006size,
  title={The size of the sync basin},
  author={Wiley, Daniel A and Strogatz, Steven H and Girvan, Michelle},
  journal={Chaos: An Interdisciplinary Journal of Nonlinear Science},
  volume={16},
  number={1},
  year={2006},
  publisher={AIP Publishing}
}

@article{calder2020calculus,
  title={The calculus of variations},
  author={Calder, Jeff},
  journal={University of Minnesota},
  volume={40},
  year={2020}
}

@article{watts1998collective,
  title={Collective dynamics of ‘small-world’networks},
  author={Watts, Duncan J and Strogatz, Steven H},
  journal={nature},
  volume={393},
  number={6684},
  pages={440--442},
  year={1998},
  publisher={Nature Publishing Group}
}

@article{panaggio2015chimera,
	title={Chimera states: coexistence of coherence and incoherence in networks of coupled oscillators},
	author={Panaggio, Mark J and Abrams, Daniel M},
	journal={Nonlinearity},
	volume={28},
	number={3},
	pages={R67},
	year={2015},
	publisher={IOP Publishing}
}

@article{omelchenko2016tweezers,
	title={Tweezers for chimeras in small networks},
	author={Omelchenko, Iryna and Omel’chenko, Oleh E and Zakharova, Anna and Wolfrum, Matthias and Sch{\"o}ll, Eckehard},
	journal={Physical Review Letters},
	volume={116},
	number={11},
	pages={114101},
	year={2016},
	publisher={APS}
}

@article{ye2025optimal,
	title={Optimal sparse networks for synchronization of semiconductor lasers},
	author={Ye, Li-Li and Vigne, Nathan and Lin, Fan-Yi and Cao, Hui and Lai, Ying-Cheng},
	journal={arXiv preprint arXiv:2511.03205},
	year={2025}
}

@article{nixon2012controlling,
	title={Controlling synchronization in large laser networks},
	author={Nixon, Micha and Fridman, Moti and Ronen, Eitan and Friesem, Asher A and Davidson, Nir and Kanter, Ido},
	journal={Physical Review Letters},
	volume={108},
	number={21},
	pages={214101},
	year={2012},
	publisher={APS}
}

@article{leyva2012explosive,
  title={Explosive first-order transition to synchrony in networked chaotic oscillators},
  author={Leyva, I and Sevilla-Escoboza, Ricardo and Buld{\'u}, JM and Sendi\~na-Nadal, Irene and G\'omez-Garde\~nes, Jesus and Arenas, Alex and Moreno, Yamir and G\'omez, Sergio and Jaimes-Reategui, Rider and Boccaletti, Stefano},
  journal={Physical Review Letters},
  volume={108},
  number={16},
  pages={168702},
  year={2012},
  publisher={APS}
}

@article{skardal2017optimal,
  title={Optimal phase synchronization in networks of phase-coherent chaotic oscillators},
  author={Skardal, Per Sebastian and Sevilla-Escoboza, Ricardo and Vera-\'Avila, VP and Buld{\'u}, Javier Mart{\'\i}n},
  journal={Chaos: An Interdisciplinary Journal of Nonlinear Science},
  volume={27},
  number={1},
  year={2017},
  publisher={AIP Publishing}
}

@article{sakaguchi1986soluble,
  title={A soluble active rotator model showing phase transitions via mutual entrainment},
  author={Sakaguchi, Hidetsugu and Kuramoto, Yoshiki},
  journal={Progress of Theoretical Physics},
  volume={76},
  number={3},
  pages={576--581},
  year={1986},
  publisher={Oxford University Press}
}

@article{filatrella2008analysis,
  title={Analysis of a power grid using a {K}uramoto-like model},
  author={Filatrella, Giovanni and Nielsen, Arne Hejde and Pedersen, Niels Falsig},
  journal={The European Physical Journal B},
  volume={61},
  number={4},
  pages={485--491},
  year={2008},
  publisher={Springer}
}

@article{matthews1990phase,
  title={Phase diagram for the collective behavior of limit-cycle oscillators},
  author={Matthews, Paul C and Strogatz, Steven H},
  journal={Physical Review Letters},
  volume={65},
  number={14},
  pages={1701},
  year={1990},
  publisher={APS}
}

@article{mikaberidze2026gradnet,
  title={GradNet: A Gradient-Based Framework for Optimal Network Science},
  author={Mikaberidze, Guram and Mikaberidze, Beso and Taylor, Dane},
  journal={arXiv preprint arXiv:2603.09197},
  year={2026}
}

@article{nikhil2025inferred,
  title={The inferred functional connectome underlying circadian synchronization in the mouse suprachiasmatic nucleus},
  author={Nikhil, KL and Singhal, Bharat and Granados-Fuentes, Daniel and Li, Jr-Shin and Kiss, Istv{\'a}n Z and Herzog, Erik D},
  journal={Proceedings of the National Academy of Sciences},
  volume={122},
  number={50},
  pages={e2520674122},
  year={2025},
  publisher={National Academy of Sciences}
}

@article{yamaguchi2003synchronization,
  title={Synchronization of cellular clocks in the suprachiasmatic nucleus},
  author={Yamaguchi, Shun and Isejima, Hiromi and Matsuo, Takuya and Okura, Ryusuke and Yagita, Kazuhiro and Kobayashi, Masaki and Okamura, Hitoshi},
  journal={Science},
  volume={302},
  number={5649},
  pages={1408--1412},
  year={2003},
  publisher={American Association for the Advancement of Science}
}

@article{onnela2011geographic,
  title={Geographic constraints on social network groups},
  author={Onnela, Jukka-Pekka and Arbesman, Samuel and Gonz{\'a}lez, Marta C and Barab{\'a}si, Albert-L{\'a}szl{\'o} and Christakis, Nicholas A},
  journal={PLOS One},
  volume={6},
  number={4},
  pages={e16939},
  year={2011},
  publisher={Public Library of Science San Francisco, USA}
}

@article{bullmore2012economy,
  title={The economy of brain network organization},
  author={Bullmore, Ed and Sporns, Olaf},
  journal={Nature Reviews Neuroscience},
  volume={13},
  number={5},
  pages={336--349},
  year={2012},
  publisher={Nature Publishing Group UK London}
}

@article{pagnier2019inertia,
  title={Inertia location and slow network modes determine disturbance propagation in large-scale power grids},
  author={Pagnier, Laurent and Jacquod, Philippe},
  journal={PloS one},
  volume={14},
  number={3},
  pages={e0213550},
  year={2019},
  publisher={Public Library of Science San Francisco, CA USA}
}

@dataset{pagnier_2022_5816149,
  author       = {Pagnier, Laurent},
  title        = {A Data Set for State and Parameter Estimation in Power Systems},
  month        = jan,
  year         = 2022,
  publisher    = {Zenodo},
  doi          = {10.5281/zenodo.5816149},
  url          = {https://doi.org/10.5281/zenodo.5816149},
}

\end{document}